\documentclass[twocolumn,prd,aps,showpacs]{revtex4}

\usepackage{amsmath}
\usepackage{verbatim}
\usepackage{graphicx}
\usepackage[usenames]{color}
\usepackage{psfrag}
\usepackage[ps2pdf,colorlinks,bookmarks]{hyperref}

\definecolor{linkblue}{rgb}{0,0,0.934}
\definecolor{linkgreen}{rgb}{0,0.60,0}
\hypersetup{pdfpagemode=None, pdfstartview=FitH, linkcolor=linkblue, %
            citecolor=linkgreen, urlcolor=linkblue}

\newcommand{\ud}[2]{^{#1}_{\phantom{#1}#2}}
\newcommand{\du}[2]{_{#1}^{\phantom{#1}#2}}

\def\beq{\begin{equation}}
\def\eeq{\end{equation}}
\def\bea{\setlength\arraycolsep{1.4pt}\begin{eqnarray}}
\def\eea{\end{eqnarray}}
\def\bit{\begin{itemize}}
\def\eit{\end{itemize}}
\def\nn{\nonumber}

\def\ie{{i.e.}}
\def\eg{{e.g.}}

\def\eq{Eq.~}
\def\eqs{Eqs.~}
\def\fig{Fig.~}

\def\pd{\partial}

\def\ld{\left}
\def\rd{\right}
\def\ra{\rightarrow}
\def\tl{\tilde}
\def\wtl{\widetilde}
\def\ph{\phantom}
\def\fr{\frac}
\def\oo{\frac{1}}
\def\half{\frac{1}{2}}
\def\const{{\rm const}}

\def\del{\delta}
\def\kap{\kappa}
\def\lam{\lambda}

\def\mn{{\mu\nu}}

\def\sig{\sigma}
\def\Sig{\Sigma}
\def\om{\omega}
\def\Om{\Omega}

\def\Rs{{}^{(3)}\!R}

\def\E{{\cal E}}

\def\O{{\cal O}}

\def\R{{\cal R}}

\def\gtrn{gauge transformation}
\def\gtrns{gauge transformations}

\def\hsf{hypersurface}
\def\hsfs{hypersurfaces}
\def\pert{perturbation}
\def\perts{perturbations}
\def\homo{homogeneous}

\def\eoms{equations of motion}

\def\bra{\langle}
\def\ket{\rangle}

\def\PR{{\cal P_{\cal R}}}
\def\Rpr{\mathcal{R}^{\rm pr}}

\begin{document}

\title{Scalar Perturbations on Lema\^itre-Tolman-Bondi Spacetimes}

\author{J. P. Zibin} \email{zibin@phas.ubc.ca}
\affiliation{Department of Physics \& Astronomy, %
University of British Columbia, %
Vancouver, BC, V6T 1Z1  Canada}

\date{\today}

\begin{abstract}

   In recent years there has been growing interest in verifying the 
horizon-scale homogeneity of the Universe that follows from applying 
the Copernican Principle to the observed isotropy.  This program has 
been stimulated by the discovery that a very large void, centred near 
us, can explain supernova luminosity distance measurements without 
dark energy.  It is crucial to confront such models with as wide a 
variety of data as possible.  With this application in mind, I develop 
the relativistic theory of linear scalar perturbations on spherically 
symmetric dust (Lema\^itre-Tolman-Bondi) spacetimes, using the covariant 
$1 + 1 + 2$ formalism.  I show that the evolution of perturbations is 
determined by a small set of new linear transfer functions.  If decaying 
modes are ignored (to be consistent with the standard inflationary 
paradigm), the standard techniques of perturbation theory on homogeneous 
backgrounds, such as harmonic expansion, can be applied, and results 
closely paralleling those of familiar cosmological perturbation theory 
can be obtained.

\end{abstract}


\maketitle

\section{Introduction}

   Since shortly after the discovery of the cosmic microwave background (CMB) 
radiation, it has been clear that the Universe is very nearly isotropic, 
with departures from uniformity in the CMB temperature at the level of 
$1$ part in $10^5$ (apart from the dipole of presumably kinetic origin).  
Since the CMB comes to us from the greatest visible distances, that observed 
isotropy tells us about the symmetry of spacetime on the largest observable 
scales.  The natural assumption is that this isotropy cannot be an 
accident of our location, and that {\em any} observer would see 
essentially the same picture, regardless of where they were located.  
This notion is known as the Copernican Principle, and implies that the 
Universe is essentially \homo\ on the largest observable scales.

   However, cosmologists are in the business of determining the nature 
of the Universe through observation and deduction, rather than through 
philosophical postulate.  As unappealing as it may appear, a Universe 
with substantial radial inhomogeneity on the largest scales is consistent 
with the observed isotropy.  Galaxy surveys are a means to quantify 
homogeneity, but are limited in their reach (see, \eg, 
\cite{hogg04,lahav02}), and {\em radial} inhomogeneity is difficult 
to disentangle from redshift-dependent effects such as evolution.  
Direct observational constraints on homogeneity over distances comparable 
to that to the last scattering surface, which emitted the CMB, are 
unavailable.

   A few novel proposals exist in the literature for observational 
signatures of radial inhomogeneity.  The CMB radiation we observe is 
partly scattered from within 
our past light cone, and this may allow us to constrain inhomogeneity 
\cite{goodman95,cs07}.  A consistency relation between luminosity 
distances and the Hubble rate, which must be satisfied for homogeneous 
and isotropic Friedmann-Robertson-Walker (FRW) models, has been discussed 
\cite{cbl07}.  Recently the time drift of cosmological redshifts was 
proposed as a test of homogeneity \cite{uce08}.

   This might appear as something of a ``dotting the i's and crossing 
the t's'' cosmological exercise, if not for the realization nearly ten 
years ago that a particular form of radial inhomogeneity can 
actually mimic the effect of accelerated expansion in supernova data 
\cite{celerier99}.  The model studied was the spherically symmetric 
dust, or Lema\^itre-Tolman-Bondi (LTB) spacetime 
\cite{lemaitre33,tolman34,bondi47}.  Since that time many studies have 
confirmed 
and elaborated upon this idea (see, \eg, \cite{aag06,em07,abnv07,gbh08}; 
more references can be found in the brief review \cite{enqvist08}).  The 
basic idea is quite simple:  the increasing expansion rate {\em in time} 
that standard dark energy models describe is replaced with increasing 
expansion rate {\em towards the centre} of a radial inhomogeneity.  
These two possibilities are difficult to distinguish if our observations 
are limited to the surface of our past light cone.

   To get the sign of the effect right (apparent acceleration rather than 
deceleration), we must be near the centre of an {\em underdensity}, or void.  
The void must be {\em nonlinear} today if it is to mimic the zeroth-order 
effect of acceleration.  Additionally, the void must extend to a great 
distance, namely several hundred Mpc, corresponding to the redshifts at which 
the relevant supernova data lies, $z \simeq 0.2 - 1$.  The presence of such 
a large void is strongly at odds with the standard picture of structure 
formation, where the largest nonlinear voids today appear at scales of tens 
of Mpc.  Nevertheless, the difficulties with understanding the theoretical 
basis for dark energy, as well as the coincidence problem, have led many 
researchers to examine such void models as alternatives to dark energy.

   In the context of void models for acceleration, and, more generally, 
considering the more fundamental issue of large-scale homogeneity, it is 
very important to constrain radial homogeneity using actual observations.  
Clearly, the more data we use to do this the better.  LTB 
models contain one (technically two; see below) free radial functions 
which could be adjusted to optimize the fit to any single data set.  
Therefore it is with 
complementary data that our greatest hope lies in constraining such 
models.  Surprizingly, few studies of void models for acceleration have 
gone beyond fitting to supernova data.  Notable exceptions are 
Refs.~\cite{aag06,abnv07}, who considered CMB data, and, very recently, 
Ref.~\cite{gbh08}, who considered a wider range of data including 
CMB and baryon acoustic oscillations.

   One important class of data which has remained out of reach so far 
involves the behaviour of scalar density \perts, \ie\ the structure we 
observe on all scales.  The evolution of this structure at {\em linear} 
order is necessarily affected by the presence of a large radial 
inhomogeneity or void in the {\em zeroth-}order background.  
While luminosity and angular diameter distances and redshifts 
are simple to calculate in LTB spacetimes, no one has yet studied 
perturbed LTB spacetimes.  It may indeed appear daunting to tackle 
\perts\ on a background spacetime which is not itself \homo\ 
\footnote{Note that LTB spacetimes are generically time-dependent, so 
it is not possible to resolve \perts\ into {\em temporal} harmonics 
as can be done on the Schwarzschild background, \eg}.  
However, with the appropriate tools, the task turns out to be not 
too much more difficult than the familiar \pert\ theory on FRW backgrounds.

   Here I develop a formalism to study the evolution of linear scalar 
\perts\ on LTB backgrounds.  The approach is fully general relativistic, 
employing the covariant $1 + 1 + 2$ formalism (see, \eg, 
\cite{greenberg70,cb03,clarkson07}), which generalizes the standard 
$1 + 3$ approach (see \cite{tcm08} for a recent review).  
This approach, while less familiar than the metric-based approach to 
many, offers several advantages, both in providing concise dynamical 
expressions which manifestly respect the symmetries of the problem, 
as well as in emphasizing physical and in principle observable 
quantities.

   I do not develop the theory of \perts\ on LTB backgrounds in 
complete generality.  Rather, I target the development towards 
application to realistic void profiles that are able to mimic 
acceleration.  My only approximation is to ignore the coupling between 
scalar and tensor modes that curved backgrounds generally facilitate.  
I argue that for the applications of interest, the sourced tensors 
will have a negligible effect on the scalar dynamics.  I determine 
only the $2$-scalar degrees of freedom, but these should be the most 
relevant observationally.  While the background is taken to be spherically 
symmetric, the \perts\ are completely free (apart from the linear and 
scalar approximations).

   It is important to choose void profiles that can fit within the 
standard inflationary paradigm, according to which at early times the 
Universe is expected to be very \homo.  [One possible origin for a 
large void (and perhaps the most conservative) is as a rare, large 
amplitude fluctuation within the primordial Gaussian random field 
produced during inflation.  The only other origin discussed for such 
a void appears to rely on highly speculative quantum cosmological 
ideas \cite{llm95}].  Therefore, such a void cannot contain a 
significant {\em decaying} mode today, without spoiling primordial 
homogeneity.  It is difficult enough to believe that such a large void 
exists and is centred near us, without compounding the difficulties 
by giving up the standard inflationary scenario as well!  Dropping the 
decaying mode also reduces our freedom to tweak an LTB model to fit 
data, so this should yield more predictive models.

   The dropping of the decaying mode provides a tremendous advantage 
in treating \perts.  Namely, since at early times the spacetime 
approaches FRW with linear fluctuations (one of which is the large 
proto-void itself), all of the familiar tools 
used in \pert\ theory on FRW can be applied at those early times, 
such as harmonic expansion and the use of statistical relations satisfied 
by primordial Gaussian random fields.  At late times, when the LTB 
background has become nonlinear, it will be possible to relate the 
scalar fluctuations to the initial conditions using a small set of 
new LTB linear transfer functions.  It should be possible to address 
many of the important observational signatures of structure using 
this formalism.  The current work lays out the formalism; 
applications to specific observations will follow \cite{voidsinprep}.

   I begin in Sec.~\ref{bgndsec} with a brief review of the relevant 
covariant $1 + 3$ formalism.  Next I describe the application of the 
further $1 + 1 + 2$ decomposition to LTB models, before describing the 
exact LTB solution.  Section~\ref{pertssec} lays out the \pert\ formalism, 
beginning with a careful definition of \perts, before introducing the 
LTB transfer functions, and then setting up the initial conditions.  
Finally, Sec.~\ref{numevsec} presents some numerical results.  I use 
signature $(-,+,+,+)$ and set $c = 1$ throughout.

\section{$1 + 1 + 2$ decomposition for LTB background}
\label{bgndsec}

   In the $1 + 3$ covariant approach to relativistic cosmology (see, \eg, 
\cite{tcm08} for a recent review), all quantities are decomposed with 
respect to a fundamental congruence of timelike worldlines, with tangent 
vector field $u^\mu$.  A spacetime 
containing a single type of matter (pressureless dust in 
the case of LTB) is perfectly suited to a covariant treatment using the 
$1 + 3$ decomposition, since the comoving worldlines naturally 
define a special timelike congruence, at least until worldlines 
intersect.  When the congruence is twist-free, it is \hsf-orthogonal 
\cite{wald84}, and hence we can define spacelike {\em slices} everywhere 
orthogonal to the congruence.

   The spherical symmetry of the LTB background is well suited 
to further decomposing matter and metric quantities according to a 
$1 + 1 + 2$ decomposition, since a spacelike congruence 
on each slice is naturally provided by the radial direction.  In this 
section I provide a covariant $1 + 1 + 2$ treatment of the LTB background, 
along the lines of that presented in Refs.~\cite{vee96,cb03,clarkson07}.  
This will set the stage for the treatment of linear \perts\ in 
Sec.~\ref{pertssec}.

\subsection{$1 + 3$ decomposition}

   To begin, we will need a set of quantities which completely describe 
the spacetime in terms of the $1 + 3$ decomposition.  Henceforth $u^\mu$ 
will always be taken to be the timelike vector field tangent to the 
comoving worldlines, with normalization $u^\mu u_\mu = -1$.  The tensor
\beq
h\ud{\mu}{\nu} \equiv \del\ud{\mu}{\nu} + u^\mu u_\nu
\label{sliceproj}
\eeq
projects orthogonal to $u^\mu$.  When the twist vanishes, we can consider 
$h_\mn$ to be the metric tensor for the orthogonal spacelike slices, and 
we can also use $h_\mn$ to define a spatially projected covariant derivative 
according to
\beq
D^{}_\mu T_{\nu_1\nu_2\cdots\nu_n}
   \equiv h\ud{\lam}{\mu} h\ud{\sig_1}{\nu_1}h\ud{\sig_2}{\nu_2}\cdots
          h\ud{\sig_n}{\nu_n} T_{\sig_1\sig_2\cdots\sig_n;\lam},
\label{slicecovder}
\eeq
for any tensor $T_{\nu_1\nu_2\cdots\nu_n}$ orthogonal to $u^\mu$ in all 
of its indices.  To describe the time evolution of any covariant quantity, 
we employ the proper time derivative along the comoving worldlines,
\beq
\dot T_{\mn\cdots\rho} \equiv u^\kap T_{\mn\cdots\rho;\kap}
\eeq
for any tensor quantity $T_{\mn\cdots\rho}$.  Finally, angled brackets 
around tensor indices indicate the spatially projected, symmetric, and 
trace-free part:
\beq
T^{\bra\mn\ket} \equiv \ld(h\ud{(\mu}{\kap}h\ud{\nu)}{\lam}
                     - \oo{3}h^\mn h_{\kap\lam}\rd) T^{\kap\lam},
\eeq
and the tensorial curl is defined by
\beq
{\rm curl}T_\mn \equiv \epsilon_{\kap\lam(\mu}D^\kap T\du{\nu)}{\lam}.
\eeq
Here $\epsilon_{\mn\lam} \equiv \epsilon_{\mn\lam\kap}u^\kap$, where 
$\epsilon_{\mn\lam\kap}$ is the totally antisymmetric four-volume element.

   We can characterize the comoving congruence kinematically via the 
following standard decomposition of the covariant derivative $u_{\mu;\nu}$:
\beq
u_{\mu;\nu} = \oo{3}\theta h_{\mn} + \sig_{\mn} + \om_{\mn}
            - a_\mu u_\nu.
\label{ucovder}
\eeq
The scalar $\theta$ measures the local volume rate of expansion of the 
congruence, while trace-free, symmetric tensor $\sig_{\mn}$ and 
antisymmetric $\om_{\mn}$ 
measure the local rates of shear and twist of the congruence, respectively.  
The vector $a_\mu \equiv u_{\mu;\rho}u^\rho$ measures the acceleration 
of the comoving worldlines.  Each of $\sig_{\mn}$, $\om_{\mn}$, and $a_\mu$ 
are orthogonal to $u^\mu$ in all of their indices.

   We will also need two tensors derived from the Weyl tensor, 
$C_{\mu\nu\lam\rho}$, which characterizes the nonlocal part of the 
gravitational field.  The electric, $E_\mn$, and magnetic, $H_\mn$, parts 
of the Weyl tensor are defined by
\beq
E_\mn \equiv C_{\mu\lam\nu\rho}u^\lam u^\rho,\quad
H_\mn \equiv \half \epsilon\du{\mu}{\lam\rho}C_{\lam\rho\nu\kappa}u^\kappa.
\eeq
The electric Weyl tensor describes the nonlocal tidal gravitational field, 
while the magnetic part describes, at least at linear order, propagating 
gravitational waves.  Both $E_\mn$ and $H_\mn$ are fully orthogonal to 
$u^\mu$.  For the case of a homogeneous and isotropic FRW cosmology, 
we have $\theta = 3H$, where $H$ is the Hubble rate, and $\sig_{\mu\nu} = 
\om_{\mu\nu} = a_\mu = E_\mn = H_\mn = 0$.

   If the pressure (and hence the anisotropic stress) vanishes, then the 
matter content is described by the simple stress-energy tensor
\beq
T^\mn = \rho u^\mu u^\nu,
\eeq
with $\rho$ the local energy density as judged by comoving observers.  The 
spatially projected stress-energy conservation law, 
$h\ud{\kap}{\nu} T\ud{\mu\nu}{;\mu} = 0$, \ie\ the momentum conservation 
law, becomes
\beq
a^\mu = 0
\eeq
(as long as $\rho \ne 0$), representing the familiar result that comoving 
dust worldlines are geodesic.

   The above kinematical, gravitational, and matter degrees of freedom 
provide a complete description of the spacetime.  It only remains to specify 
their evolution.  The twist $\om_\mn$ of the congruence is a special case, 
and satisfies the evolution equation \cite{tcm08}
\beq
h\ud{\mu}{\nu}\dot\om^\nu = -\fr{2}{3}\theta\om^\mu + \sig\ud{\mu}{\nu}\om^\nu,
\eeq
where $\om_\mu \equiv \epsilon_{\mn\lam}\om^{\nu\lam}/2$.  Therefore, if 
the twist vanishes initially, it vanishes for all time.  Henceforth the 
twist will always be set to zero.  This amounts to ignoring the presence 
of vector perturbation modes at linear order, which is a reasonable 
assumption within the standard inflationary scenario.  Also, this means 
we can define spacelike slices orthogonal to the comoving flow.

   The evolution of the remaining quantities is determined by the following 
set of equations \cite{tcm08}:\begin{widetext}
\bea
\textrm{Energy conservation:}& \qquad &\dot\rho = -\theta\rho\ph{\fr{2}{3}}
\label{encons}\\
\textrm{Raychaudhuri:}& \qquad & \dot\theta = -\fr{\theta^2}{3}
             - 4\pi G\rho - \sig^{\mu\nu}\sig_{\mu\nu}
\label{Raychaud}\\
\textrm{Shear evolution:}& \qquad & \dot\sig_{\bra\mn\ket}
                = -\fr{2}{3}\theta\sig_\mn
                - \sig_{\lam\bra\mu}\sig\ud{\lam}{\nu\ket} - E_\mn
\label{Sigevoln}\\
E_\mn \textrm{ evolution:}& \qquad & \dot E_{\bra\mn\ket} = -\theta E_\mn
           + 3\sig_{\lam\bra\mu}E\ud{\lam}{\nu\ket}
           - 4\pi G\rho\sig_\mn + {\rm curl}H_\mn\ph{\fr{2}{3}}
\label{Eevoln}\\
H_\mn \textrm{ evolution:}& \qquad & \dot H_{\bra\mn\ket} = -\theta H_\mn 
     + 3\sig_{\lam\bra\mu}H\ud{\lam}{\nu\ket} - {\rm curl}E_\mn.\ph{\fr{2}{3}}
\label{Hevoln}
\eea
\end{widetext}The evolution of each quantity is determined 
by a damping term, proportional to the expansion (possibly corrected by 
the shear), together with coupling terms, such as the direct coupling 
between shear and the tidal field $E_\mn$, and the coupling between 
electric and magnetic Weyl curvature via the curl terms, which supports the 
propagation of gravitational waves.  This set of equations would have been 
signficantly more complicated had we not chosen $u^\mu$ to be comoving, 
with extra terms involving the momentum density and acceleration $a^\mu$.  
In addition, the initial conditions must satisfy a set of initial-value 
constraint equations, although we will not need the explicit form of the 
constraints here.

\subsection{$1 + 1 + 2$ decomposition}

   Under spherical symmetry, each comoving-orthogonal slice contains a 
preferred spacelike congruence with radial tangent vector $r^\mu$, where 
$r^\mu r_\mu = 1$.  By analogy with the tensor $h\ud{\mu}{\nu}$ defined 
in \eq(\ref{sliceproj}), which projects into the slices, we can define a 
tensor $s_\mn$ by
\beq
s_\mn \equiv h_\mn - r_\mu r_\nu,
\label{sheetproj}
\eeq
which projects orthogonally to both $u^\mu$ and $r^\mu$.  A $2$-surface 
lying in a slice and orthogonal to $r^\mu$ will be called a {\em sheet.}  
(The sheets are ordinary $2$-spheres under spherical symmetry.)

    Any symmetric, trace-free tensor $T_\mn$ which is orthogonal to $u^\mu$ 
(\ie\ which is a $3$-{\em tensor}) can be decomposed according to
\beq
T_\mn = {\cal T}\ld(r_\mu r_\nu - \half s_\mn\rd) + 2{\cal T}_{(\mu}r_{\nu)}
      + {\cal T}_\mn.
\label{sheetdecomp}
\eeq
Here ${\cal T} \equiv r^\mu r^\nu T_\mn$ is a fully radially projected 
$2$-scalar, ${\cal T}^{\mu} \equiv s\ud{\mu}{\nu}r_\lam T^{\nu\lam}$ is 
a projected $2$-vector, and ${\cal T}^\mn \equiv 
\ld(s\ud{(\mu}{\kap}s\ud{\nu)}{\lam} - s^\mn s_{\kap\lam}/2\rd) T^{\kap\lam}$ 
is a symmetric, trace-free $2$-tensor.  Any $3$-vector can be similarly 
decomposed, although we will not need the explicit result.  Any $3$-scalar 
such as the energy density $\rho$ is automatically also a $2$-scalar.

   The utility of this $1 + 1 + 2$ decomposition becomes immediately 
apparent when we realize that, under spherical symmetry, all $2$-vectors 
and $2$-tensors must vanish.  Then, with straightforward manipulations, 
the evolution \eqs(\ref{encons}) to (\ref{Eevoln}) become
\beq
\dot\rho = -\theta\rho,
\label{enconsscal}
\eeq
\beq
\dot\theta = -\fr{\theta^2}{3} - 4\pi G\rho - \fr{3}{2}\Sig^2,
\eeq
\beq
\dot\Sig = -\ld(\fr{2}{3}\theta + \half\Sig\rd)\Sig - \E,
\eeq
\beq
\dot \E = -\ld(\theta - \fr{3}{2}\Sig\rd)\E - 4\pi G\rho\Sig.
\label{Eevolnscal}
\eeq
Here I have defined the $2$-scalars $\Sig \equiv r^\mu r^\nu\sig_\mn$ 
and $\E \equiv r^\mu r^\nu E_\mn$.  The set of tensorial partial 
differential equations, (\ref{encons}) to (\ref{Hevoln}), has been 
replaced with a simple, closed set of scalar ordinary differential 
equations (ODEs).  The magnetic Weyl tensor vanishes under spherical 
symmetry: no gravitational waves are possible in this case.

   An equivalent way to see that all quantities reduce to $2$-scalars 
in the spherically symmetric case is that any $3$-tensor or $3$-vector 
can only be constructed from $r^\mu$ and $s^\mn$; no other objects are 
available.  [\eq(\ref{sheetproj}) relates $h_\mn$ to $r_\mu$ and $s_\mn$.]  
Thus, \eg, the trace-free shear tensor $\sig_\mn$ must take the form
\beq
\sig_\mn = \ld(r_\mu r_\nu - \half s_\mn\rd)\Sig.
\eeq

\subsection{Exact solution}
\label{exactsolnsec}

   The set of ODEs, \eqs(\ref{enconsscal}) to (\ref{Eevolnscal}), could 
be numerically integrated, given initial conditions that satisfy the 
constraint equations.  However, as is well known, an exact solution 
exists for the spherically symmetric dust spacetime 
\cite{lemaitre33,tolman34,bondi47}.  
It is normally expressed in terms of the metric, but can be readily 
rewritten in terms of the covariant $2$-scalars $\rho$, $\theta$, $\Sig$, 
and $\E$.  To express the exact solution, we must first introduce a time 
coordinate $t$ that labels the comoving-orthogonal slices.  Because 
the comoving worldlines are geodesic, we can choose $t$ to measure proper 
time along the worldlines; henceforth this choice will always be assumed.  
Similarly, it is useful to define a radial coordinate, $r$, which is 
constant along the comoving worldlines.

   The result is
\beq
4\pi G\rho = \fr{M'}{Y^2Y'},
\label{rhoexact}
\eeq
\beq
\theta = \fr{\dot Y'}{Y'} + \fr{2\dot Y}{Y},
\eeq
\beq
\Sig = \fr{2}{3}\ld(\fr{\dot Y'}{Y'} - \fr{\dot Y}{Y}\rd),
\eeq
\beq
\E = \fr{8\pi G}{3}\rho - \fr{2M}{Y^3},
\label{Eexact}
\eeq
where $Y = Y(r,t)$ is given implicitly by
\beq
Y = \fr{M}{K}(1 - \cosh\eta),\quad
t - t_B = \fr{M}{(-K)^{3/2}}(\sinh\eta - \eta),
\label{Ysoln}
\eeq
and $M = M(r)$, $K = K(r) < 0$, and $t_B = t_B(r)$ are arbitrary radial 
functions.  Here the prime symbol denotes $\partial/\partial r$.  
(Related solutions are known for the cases $K \ge 0$, but 
will not be needed here.)  These expressions for the $2$-scalars were 
first written down in \cite{st02}.  
Given values of $t$ and $r$ for which a solution is desired, and choices 
for the three arbitrary functions $M(r)$, $K(r)$, and $t_B(r)$, we can 
solve \eq(\ref{Ysoln}) for $Y(r,t)$ and its derivatives.  Then we can 
evaluate each of \eqs(\ref{rhoexact}) to (\ref{Eexact}) to complete the 
solution.  It is straightforward to verify explicitly that 
\eqs(\ref{rhoexact}) to (\ref{Ysoln}) solve the ODEs (\ref{enconsscal}) 
to (\ref{Eevolnscal}) (as well as the initial-value constraints) for 
arbitrary $M(r)$, $K(r)$, and $t_B(r)$.

   Because we are free to reparametrize the radial coordinate $r$, only 
{\em two} of the radial functions $M(r)$, $K(r)$, and $t_B(r)$ describe 
physically distinct solutions.  Therefore, the general spherically symmetric 
dust solution depends on two free radial functions.  It will be crucial 
to understand the physical significance of these two functions, which 
is the subject of the 
\hyperref[growdecayapp]{appendix}, but we can see readily from 
the above exact solution that as $t \ra t_B(r)$, for fixed $r$, the density 
$\rho \ra \infty$, \ie\ we approach a physical singularity.  This 
singularity is analogous to the initial singularity in the homogeneous and 
isotropic FRW spacetime, but with extra radial dependence; hence the name 
``bang time'' for the function $t_B(r)$.

   In the \hyperref[growdecayapp]{appendix}, it is shown that the two 
free radial LTB functions correspond to the growing and decaying mode for 
linear \perts\ about FRW backgrounds.  The decaying mode corresponds to 
fluctuations in the bang time function, $t_B(r)$, and can be set to zero 
by choosing $t_B(r) = \const$.  The growing mode corresponds to fluctuations 
in the spatial Ricci curvature $\Rs$ of the comoving slices, and can be 
made to vanish with the choice $\Rs = 0$.  It will be very important to 
eliminate the decaying mode in setting up realistic void profiles, 
since otherwise at early times the Universe becomes 
extremely inhomogeneous, contradicting the standard picture of extreme 
post-inflationary homogeneity.

   Although it will not be needed in this work, to facilitate comparison 
with previous studies the LTB metric can be written explicitly as
\beq
ds^2 = -dt^2 + \fr{Y'^2}{1 - K}dr^2 + Y^2d\Omega^2.
\eeq
It reduces to the FRW metric in the homogeneous case (where the shear and 
electric Weyl scalars, $\Sig$ and $\E$, both vanish), and to the 
Schwarzschild metric outside some radius $r = L$, if $\rho(r) = 0$ for 
$r > L$.

\section{Scalar \perts\ about LTB backgrounds}
\label{pertssec}
\subsection{Scalar-tensor coupling}
\label{scaltenssec}

   The problem of evolving linear \perts\ on a spherically symmetric dust 
(LTB) background is in principle straightforward:  we must linearize the 
set of evolution equations (\ref{encons}) to (\ref{Hevoln}) about an 
arbitrary LTB background solution.  In doing so, the symmetry of the 
background implies that any $2$-vector or $2$-tensor quantities must be 
of first order, and hence their products can be ignored.

   First consider a $2$-scalar subset of the full set, namely 
\eqs(\ref{encons}) and (\ref{Raychaud}) and the contraction between 
$r^\mu r^\nu$ and \eqs(\ref{Sigevoln}) and (\ref{Eevoln}).  Since we must 
relax (at linear order) the condition of spherical symmetry, the vector 
field $r^\mu$ can no longer be chosen to be exactly radial.  Nevertheless, 
we can suppose that $r^\mu$ departs from radial only at first order.  We 
will see in Sec.~\ref{pertdefnsec} that $2$-scalar quantities are 
invariant under first-order variations in the field $r^\mu$, so we have 
no need to fix the field at first order.  Straightforward manipulations 
then lead to the $2$-scalar linearized set:
\beq
\dot\rho = -\theta\rho,
\label{enconslin}
\eeq
\beq
\dot\theta = -\fr{\theta^2}{3} - 4\pi G\rho - \fr{3}{2}\Sig^2,
\label{Raychaudlin}
\eeq
\beq
\dot\Sig = -\ld(\fr{2}{3}\theta + \half\Sig\rd)\Sig - \E,
\eeq
\beq
\dot \E = -\ld(\theta - \fr{3}{2}\Sig\rd)\E - 4\pi G\rho\Sig
        + \epsilon_\mn\nabla^\mu\cal H^\nu.
\label{Eevolnlin}
\eeq
Here $\cal H^\nu$ is the $2$-vector part of the $1 + 1 + 2$ decomposition 
of $H_\mn$; recall \eq(\ref{sheetdecomp}).  The symbol $\nabla^\mu$ 
represents the sheet-projected covariant derivative, defined in analogy 
with \eq(\ref{slicecovder}), but with sheet projection tensors $s_\mn$ 
replacing $h_\mn$; \ie, it represents the covariant derivative in the 
``angular'' directions.  Finally, $\epsilon_\mn \equiv 
\epsilon_{\mn\lam}r^\lam$.  This set of equations agrees with the 
corresponding linearized subset of the full set derived in \cite{clarkson07}.  
Again, a set of initial-value constraint equations is not shown.  

   Remarkably, this set of $2$-scalar equations is identical to the 
corresponding exact set under spherical symmetry, \eqs(\ref{enconsscal}) 
to (\ref{Eevolnscal}), with the exception of the term 
$\epsilon_\mn\nabla^\mu\cal H^\nu$ which couples $\E$ to the magnetic Weyl 
tensor.  Physically, this term couples scalar \pert\ modes to tensor 
modes, represented by $H_\mn$ \footnote{Technically, we expect the 
tensors to split into even and odd parity modes.  Only the even parity 
tensor modes couple to the scalar modes \cite{clarkson07}.}.  While it 
is well known that scalars and 
tensors decouple when linearizing about FRW, such a coupling is expected 
to be a feature of the evolution on more general backgrounds, since a 
nontrivial background allows one to construct linear scalar quantities 
via, \eg, contractions between the background and tensor modes.  (Recall 
as well that scalars couple to tensors at {\em second} order about FRW 
backgrounds, essentially since the second-order variables propagate in a 
non\homo\ first-order background.)

   Unfortunately, because of the scalar-tensor coupling, the 
linearized set, (\ref{enconslin}) to (\ref{Eevolnlin}), 
is not closed, and to complete the description of the dynamics we must 
include the $2$-vector part of the $H_\mn$ evolution equation, 
\eq(\ref{Hevoln}).  This equation in turn couples to $2$-vector and 
$2$-tensor parts of the shear and electric and magnetic Weyl tensors, 
resulting in a very large set of equations \cite{clarkson07}.  On top 
of the added complexity of several more equations, the coupling to 
$H_\mn$ means that we must evolve a set of {\em partial} differential 
equations.  If the coupling were absent, we would only need to solve 
a much simpler set of {\em ordinary} differential equations.  Indeed, 
as I mentioned above, 
in that case the set of equations would be {\em identical} to the 
corresponding set under spherical symmetry, and hence we could employ 
the known exact solution to those equations!  (Of course, since we 
have ignored second-order terms in \eqs(\ref{enconslin}) to 
(\ref{Eevolnlin}), the use of the exact solution in this way could only 
be trusted up to first order.)

   The vast simplification of the dynamics in the absence of the tensor 
coupling leads naturally to the question:  Under what conditions 
could the tensor coupling be ignored to a good approximation?  It is 
certainly reasonable to consider the case where tensors are negligible 
at {\em early} times, before the LTB background becomes nonlinear (\eg\ 
at the time of last scattering), since many viable inflation models 
predict a very small primordial tensor contribution.  However, even in 
this case, tensor modes will generally be sourced by scalars at late 
times, when the LTB background becomes nonlinear, according to 
\eq(\ref{Hevoln}).

   Nevertheless, in the case of interest, namely that of linear scalar 
\perts\ on top of a mildly nonlinear spherical inhomogeneity, it is not 
expected that significant tensors will be sourced.  It {\em is} expected 
that tensors would be sourced at times late enough, or scales small 
enough, that the scalar \perts\ {\em themselves} have become nonlinear 
\cite{cbm07}.  Essentially, nonlinear collapsing overdensities during 
structure formation are generically nonspherical, and hence source 
tensors through a significant time-dependent quadrupole moment.  
Instead, in the case of interest in the present work, we have 
{\em linear} fluctuations on top of a nonlinear background which itself 
cannot source tensor modes.

   In Ref.~\cite{mhm97}, the authors performed a numerical study of the 
linear stability of $H_\mn = 0$ in a few exact backgrounds, including a 
planar inhomogeneous Szekeres model.  They found that vanishing magnetic 
Weyl curvature was stable during pancake collapse, but unstable during 
collapse 
towards a spindle geometry.  This supports the claim that tensors can 
be ignored in the present context, where in the late stages of evolution 
of a void, the outer regions tend to collapse to form an overdense shell, 
which locally approximates pancake collapse geometry.

   Note that dynamics with vanishing pressure and magnetic Weyl curvature 
are generally 
described by a set of ODEs along the comoving worldlines, as can be 
seen directly from the exact set, \eqs(\ref{encons}) to (\ref{Eevoln}).  
This was noted some time ago \cite{mps93}, and such a spacetime was 
dubbed {\em silent}.  This name refers to the fact that without pressure 
and $H_\mn$, sound waves and gravitational waves cannot be supported, 
respectively, and hence no direct communication can exist between 
neighbouring worldlines.  Apart from a few special cases, such spacetimes 
are thought to be generally inconsistent at the nonlinear level, in that 
the condition $H_\mn = 0$ imposes a new constraint on the dynamics which 
is not consistent with the remaining dynamical equations \cite{veulem97}.  
However, in the present context we are only assuming that the coupling to 
tensors can be ignored at linear order.  Also, we do not require that 
the full tensor $H_\mn$ vanish, only the weaker condition that 
$\epsilon_\mn\nabla^\mu\cal H^\nu$ can be ignored.

   Henceforth I will assume that the tensor coupling term can be ignored 
in \eq(\ref{Eevolnlin}) in evolving \perts\ on LTB backgrounds.  
As a consistency check, I show in Section~\ref{numevsec} that for 
a particular LTB profile chosen to provide a rough fit to the luminosity 
distance-redshift relation of standard $\Lambda$ cold dark matter 
($\Lambda$CDM) models, the background shear, through which tensors 
must couple to density \perts, only very weakly effects the density \perts.  
Such a test should be 
performed for any such calculation of scalar \perts\ on LTB backgrounds.  
Obviously the full set of equations, including tensors, could in principle 
be evolved to check the validity of the weak tensor-scalar coupling 
approximation.  This might be necessary if we were interested in evolving 
\perts\ on spherical {\em overdensities} at the late stages of collapse, 
instead of underdensities.

\subsection{Defining \perts}
\label{pertdefnsec}

   The dynamical equations derived in Section~\ref{scaltenssec} describe 
the evolution of the {\em total} quantities $\rho$, $\theta$, $\Sig$, 
and $\E$, \ie\ backgrounds plus \perts, in the linear approximation.  
For many purposes it is very useful to obtain the evolution of appropriately 
defined \perts\ alone.  To do this in the metric-based approach to \pert\ 
theory, one first provides a mapping (\ie\ makes a gauge choice) between 
the real spacetime and a fictitious background (usually FRW).  
Perturbations in any quantity are then defined as the difference between 
the exact value at some event and the background value mapped to that 
event.  The freedom to vary the mapping results in the familiar gauge 
ambiguity.

   On the other hand, within the $1 + 3$ covariant approach to \pert\ 
theory (see \cite{tcm08} for a recent review) \perts\ are usually 
represented by spatial {\em gradients} orthogonal to the timelike 
direction $u^\mu$; \eg, 
density fluctuations are characterized by the $3$-vector $D_\mu\rho$.  
The intention is to describe fluctuations in a gauge-invariant and 
covariant manner.  However, it is easy to see that precisely the same 
ambiguity exists in this approach as in the metric approach.  If we 
choose $u^\mu$ to be orthogonal to \hsfs\ of constant energy density, 
then we have $D_\mu\rho = 0$ trivially, \ie\ we have ``gauged away'' 
the density \pert.  Of course, as we have seen here, in the case of a 
single-component matter source, there is a natural choice for $u^\mu$ 
which simplifies the dynamical equations, namely the comoving choice.  
But this choice is not necessary.

   While the ability to define \perts\ entirely within the true spacetime is 
an advantage of the covariant approach for FRW backgrounds, it does not 
appear to be possible to extend this approach to LTB backgrounds.  This is 
because, on the natural comoving-orthogonal slices, gradients such as 
$D_\mu\rho$ do not generally vanish at {\em background} level, and hence 
they cannot be said to characterize the \perts\ alone.  Therefore, here I 
will employ a more standard approach, defining the density \pert\ as 
the difference
\beq
\del\rho(x^\mu) \equiv \rho(x^\mu) - \rho^{(0)}(r, t).
\label{drhodef}
\eeq
Here $\rho(x^\mu)$ is the density at event $x^\mu$ in the true, perturbed 
spacetime, and $\rho^{(0)}(r, t)$ is the density at the corresponding event 
in the background, in this case a spherically symmetric LTB spacetime.  
I define \perts\ in the $2$-scalars $\theta$, $\Sig$, and $\E$ analogously.  
The comoving worldlines in a dust spacetime provide a natural choice for 
the time coordinate.  This coordinate is chosen (in both true and 
background spacetimes) such that \hsfs\ of constant $t$ are orthogonal 
to the comoving worldlines, and $t$ measures proper time along those 
worldlines.  This fixes the ``temporal gauge.''

   There is inevitable freedom in choosing the radial coordinate $r$, 
since no radial vector $r^\mu$ is naturally defined in the perturbed 
spacetime, and there is no unique notion of distance in a curved spacetime.  
We can defeat this ambiguity by considering the intended application of 
the perturbed LTB formalism, namely the evolution of linear scalar 
\perts\ on top of a spherical void.  Such a void becomes nonlinear at 
late times, but, as discussed in the \hyperref[growdecayapp]{appendix}, 
must consist of a pure growing mode, to be consistent with the 
inflationary paradigm.  Hence the spacetime approaches FRW plus linear 
\perts\ at early times (such as the time of last scattering).  We can 
define $r$ in the FRW background at such early times to give radial 
proper distance from some centre.  Then we can define $r$ on the 
perturbed spacetime at early times to be {\em any} radial coordinate that 
agrees, to first order, with the background radial coordinate.  Taking 
the background at this early time to be \homo\ FRW, any {\em linear} 
variations $r \ra r + \del r$ in the radial coordinate {\em cannot} 
affect the values of any perturbed physical quantities, to first order.  
({\em Physical} \perts\ in FRW are invariant under purely {\em spatial} 
\gtrns, at first order \cite{bardeen88,hwang91}.)  
We then define $r$ at later events such that it is constant 
along each comoving worldline in the true spacetime, \ie\ $r$ is 
``dragged'' along by the fluid flow.  (Therefore comoving worldlines 
in the background are mapped to comoving worldlines in the perturbed 
spacetime.)  The spacelike vector field $r^\mu$ 
is defined to be normal to surfaces of constant $r$.  These surfaces 
will be linearly perturbed versions of $2$-spheres.

   Note that a type of radial gauge dependence may be present in \perts\ 
at late times.  In regions where the background energy denstity, \eg, 
has a significant spatial gradient, it may be possible to gauge away 
a density \pert\ $\del\rho$ with a radial \gtrn.  (Indeed this gauge 
dependence may continue to early times if we consider the background 
to be LTB, rather than FRW, at the early times.)  However, we will not 
be able to gauge away $\del\rho$ everywhere, since the density gradient 
will not be large everywhere ($d\rho/dr$ must vanish at the origin if 
$\rho$ is to be smooth there).  In the FRW case, the usual monotonicity 
of the background $\rho(t)$ does allow us to set $\del\rho = 0$ everywhere.  
The possible gauge dependence in the LTB case must be kept in mind when 
interpreting the results of calculations.  Of course, a calculation of 
any {\em observable} quantity must be independent of any gauge choice.  
[An example would be the variance of $\del\rho$ across a $2$-sphere of 
constant redshift (which is directly observable), rather than some 
arbitrary coordinate $r$ (which is not).]

   Recall that in the $1 + 1 + 2$ formalism, $2$-scalars are constructed 
from $3$-vectors or $3$-tensors by 
fully projecting along $r^\mu$.  Thus we might be concerned that linear 
variations in $r^\mu$, which will necessarily follow from linear 
variations in the coordinate $r$, might change the values of $2$-scalars, 
at least at late times when the background has become nonlinear.  
However, this does not happen, {\em at first order,} because of the 
isotropy of the background.  Consider an exact, spherically symmetric 
LTB background, with exactly radial vector $r^\mu$ and linearly perturbed 
vector
\beq
\tl r^\mu \equiv r^\mu + \del r^\mu.
\eeq
Imposing the normalization $\tl r^\mu \tl r_\mu = 1$ gives 
$\tl r^\mu r_\mu = 1 + \O(2)$.  Therefore, defining the shear scalar 
$\Sig$ using the radial $r^\mu$, $\Sig \equiv r^\mu r^\nu\sig_\mn$, we have
\beq
\wtl\Sig \equiv \tl r^\mu \tl r^\nu \sig_\mn = \Sig + \O(2).
\eeq
(A similar result applies to the electric Weyl tensor.)  That is, linear 
variations in $r^\mu$ do not affect the values of 
$2$-scalars, to first order.  Clearly the presence of \perts\ will not 
change this result, since the variations in the values of $2$-scalars 
due to the presence of linear \perts\ as $r^\mu$ is varied at first order 
will also be of at least second order.

   To summarize, the temporal gauge is chosen to be comoving, while the 
radial gauge is any linear variation about ``initial slice proper distance 
gauge.''  The temporal choice simplifies the dynamical equations, while 
the radial choice enables the simple adaptation of standard results from 
the theory of \perts\ on FRW backgrounds, as we will see in 
Sec.~\ref{harmdecompsec}.

\subsection{LTB transfer functions}
\label{transfncsec}

   Inserting the definition of $\del\rho$, \eq(\ref{drhodef}), and the 
corresponding expressions for the other $2$-scalar \perts, into the 
linearized set, \eqs(\ref{enconslin}) to (\ref{Eevolnlin}) (dropping the 
tensor coupling), and discarding terms quadratic in the \perts, we obtain
\beq
\del\dot\rho = -\theta\del\rho - \rho\del\theta,
\label{denconslin}
\eeq
\beq
\del\dot\theta = -\fr{2}{3}\theta\del\theta - 4\pi G\del\rho - 3\Sig\del\Sig,
\label{dRaychaudlin}
\eeq
\beq
\del\dot\Sig = -\ld(\fr{2}{3}\theta + \Sig\rd)\del\Sig
             - \fr{2}{3}\Sig\del\theta - \del\E,
\label{dSigevolnlin}
\eeq
\bea
\del\dot\E = &-&\ld(\theta - \fr{3}{2}\Sig\rd)\del\E
           - \E\ld(\del\theta - \fr{3}{2}\del\Sig\rd)\nn\\
           &-& 4\pi G(\rho\del\Sig + \Sig\del\rho).
\label{dEevolnlin}
\eea
In obtaining these equations, the fact that comoving worldlines in the 
background are mapped to comoving worldlines in the perturbed spacetime 
has allowed us to subtract off the background \eoms, \eqs(\ref{enconsscal}) 
to (\ref{Eevolnscal}).  Also, note that it is valid to trivially perturb time 
derivatives (\eg, $\del\dot\rho = \dot\rho - \dot\rho^{(0)}$), since $t$ 
measures proper time in both background and perturbed spacetimes.

   In a sense, this completes the problem of \perts\ on LTB backgrounds, 
since the set of ODEs, \eqs(\ref{denconslin}) to (\ref{dEevolnlin}), can 
be readily solved numerically (or even using the exact solution from 
Sec.~\ref{exactsolnsec}, remembering that that solution can only 
be trusted to linear order here).  However, with applications in mind, 
it will be very useful to develop a description of the evolution of the 
\perts\ in terms of a new set of {\em transfer functions.}  This will 
allow us to derive results that parallel remarkably closely the standard 
results from \pert\ theory on FRW.

   Consider the column vector of $2$-vector \perts, $\del X_i(t,r,n^\mu)$, 
where $X_i = \rho$, $\theta$, $\Sig$, $\E$.  Here $n^\mu$ is the spatial 
unit vector from the origin, $r = 0$, towards the point at which $\del X_i$ 
is evaluated, so $(r,n^\mu)$ are the polar coordinates of that point.  Then 
the linear \perts\ at some arbitrary time $t$ must be related to those at 
some initial time $t_i$ via the matrix product
\beq
\del X_i(t,r,n^\mu) = T_{ij}(t,r) \del X_j(t_i,r,n^\mu),
\label{gentransfnc}
\eeq
where $T_{ij}(t,r)$ is a $4\times4$ matrix of transfer functions.  
Linearity of evolution demands that \eq(\ref{gentransfnc}) take a linear 
form.  The property of ``silence,'' \ie\ that the \perts\ are solutions 
of ODEs, means that the solution at event $(t,r,n^\mu)$ can only 
depend on the initial condition on the same worldline at 
$(t_i,r,n^\mu)$.  (Initial velocities $\del\dot X_i$ are not 
needed since the ODEs are first order.)  Finally, the isotropy of the 
background implies that the transfer functions $T_{ij}$ must depend 
only on $t$ and $r$.

   In the case of \perts\ on FRW, the corresponding transfer functions 
are generally scale (\ie\ $k$) dependent, although this is only true 
when pressure is present to support acoustic waves and introduce 
spatial derivatives into the dynamical equations.  In the LTB case 
here, no such scale dependence exists.  Also, the FRW transfer functions 
appear to be considerably simpler than their LTB counterparts, in that they 
depend on $t$ alone in the \homo\ case.  However, again with applications 
in mind, we will very often be interested in evaluating the \perts\ on 
the observer's past light cone (with the observer located at the centre 
of symmetry).  The surface of that cone defines a definite relationship 
$r = r(t)$ (at background order), and so the LTB transfer functions can 
in practice be considered to be functions of the time alone, 
$T_{ij}(t,r) = T_{ij}(t,r(t))$.

   Thus the problem of linear \perts\ on LTB backgrounds, which at first 
may have appeared quite daunting, is reduced to that of calculating a 
small number of transfer functions of a single variable, for most 
practical applications.  As we will see next, the problem can be simplified 
even further by consideration of the initial conditions.

\subsection{Initial conditions}

   In order to use \eq(\ref{gentransfnc}) to evolve the perturbations we 
must first specify initial conditions.  To do this, first recall again 
the context in which we are considering LTB backgrounds here.  We require 
growing mode solutions, which approach linearly perturbed FRW at early 
times.  Therefore, we can set up initial conditions for the fluctuations 
at some sufficiently early time $t_i$, exactly as in the standard FRW case.  
For definiteness (and well supported by current observations \cite{wmap5}), 
I will assume that the primordial fluctuations are scalar and adiabatic.  
In addition, with their inflationary origin in mind, I will take 
the \perts\ at the initial time, $t_i$, to be purely in the growing mode 
(this assumption is also supported by the observations \cite{af05}).  
Since the above calculations have assumed vanishing pressure, we must 
take the initial time for the specification of \perts\ to be sufficiently 
late that matter domination is a good approximation.

   With these assumptions, we can reduce the number of LTB transfer 
functions that need to be calculated, since some of the initial \perts\ 
can be written in terms of the others.  It will be particularly useful 
to write all of the initial \perts\ in terms of the comoving curvature 
\pert, $\R(x^\mu)$.  This quantity is constant on super-Hubble scales (for 
adiabatic modes), and hence is very convenient for comparing with the 
predictions of a particular inflationary model.  Constraints on the 
primordial value of $\R$ are also explicitly provided in observational 
results (see, \eg, \cite{wmap5}).

   During matter domination, the comoving energy density \pert\ can be 
written in terms of $\R$ as
\beq
\fr{\del\rho}{\rho} = -\fr{18}{5}\fr{D^2}{\theta^2}\R
\label{rholinIC}
\eeq
(see, \eg, \cite{ll00}), where $D^2 \equiv D^\mu D_\mu$ is the spatial 
Laplacian.  At the initial time $t_i$, the shear tensor for scalar 
fluctuations can be derived from a scalar function $\sig(x^\mu)$, according to
\beq
\sig_\mn = D_\mu D_\nu \sig - \oo{3}h_\mn D^2\sig.
\eeq
Again during matter domination, the comoving shear scalar $\sig$ can be 
written in terms of the comoving curvature \pert\ as
\beq
\sig = \fr{6}{5}\fr{\R}{\theta}
\eeq
(see, \eg, \cite{hwang91}).  Combining these results we can write for the 
shear $2$-scalar initial condition at $t_i$,
\beq
\fr{\del\Sig}{\theta} = \fr{r^\mu r^\nu\sig_\mn}{\theta}
   = \fr{6}{5}\oo{\theta^2}\ld(\fr{\pd^2}{\pd r^2} - \oo{3}D^2\rd)\R.
\label{shearlinIC}
\eeq

   Note that, even though both $\del\rho$ and $\del\Sig$ are completely 
determined by the comoving curvature $\R$, they are not simply 
proportional.  Geometrically, it is possible, \eg, that some localized 
linear fluctuation in $\R$ produces a $\del\rho$ at some point, but 
no $\del\Sig$, if the fluctuation is symmetric about the point.  
Nevertheless, in the \hyperref[growdecayapp]{appendix}, \eqs(\ref{delrhoFRW}) 
and (\ref{delthetaFRW}) show that the growing modes for the comoving gauge 
energy density and expansion \perts\ are related by
\beq
\fr{\del\rho}{\rho} = -3\fr{\del\theta}{\theta}.
\label{rhoproptheta}
\eeq
Similarly, using \eq(\ref{dSigevolnlin}) (for the relevant case at $t_i$ 
that the background shear $\Sig$ is first order and hence can be dropped 
from this equation) and \eq(\ref{shearlinIC}), we can relate the electric 
Weyl $2$-scalar to the shear via
\beq
\del\E = -\half\theta\del\Sig.
\label{EpropSig}
\eeq

   It is important to point out that these initial conditions must 
automatically satisfy the Einstein initial-value constraint equations, to 
linear order.  Hence all that remains is to determine the transfer 
functions to evolve the initial conditions.

   The proportionalities (\ref{rhoproptheta}) and (\ref{EpropSig}) apply 
at the initial time, $t_i$, and hence they reduce the dimensionality of 
the matrix of transfer functions in \eq(\ref{gentransfnc}) to $2\times2$.  
For example, the evolution of the density \pert\ is given by
\bea
\fr{\del\rho(t,r,n^\mu)}{\rho(t,r)}
   &=& T_{\rho\rho}(t,r)\fr{\del\rho(t_i,r,n^\mu)}{\rho(t_i)}\nn\\
   &-& T_{\rho\Sig}(t,r)\fr{3\del\Sig(t_i,r,n^\mu)}{\theta(t_i)},
\label{delrhotransfnc}
\eea
and we only need to calculate two transfer functions.  In this expression 
I have scaled the \perts\ so that the transfer functions are dimensionless 
and positive, and the function $T_{\rho\rho}$ approaches unity at early times 
($T_{\rho\Sig}$ approaches zero at early times).

\subsection{Harmonic decomposition}
\label{harmdecompsec}

   One of the difficulties that comes to mind when first considering the 
problem of linear \perts\ on LTB backgrounds is that of harmonic 
expansion.  A crucially important tool in \pert\ theory on flat FRW 
backgrounds is the ability to expand fluctuations in Fourier modes 
or spherical harmonics and spherical Bessel functions.  This approach 
can be generalized to the case of spatially curved FRW backgrounds.  
But an LTB background can have an {\em arbitrary} radial dependence 
of the spatial curvature, so it appears hopeless to look for radial 
harmonics generally (although the standard spherical harmonic functions, 
the $Y_{\ell m}$'s, can be used to expand the angular dependence, of 
course).

   Again, though, we are considering spacetimes which approach 
linearly perturbed FRW at early times.  Therefore we should be able 
to employ the usual harmonic expansions {\em at a sufficiently early 
time.}  Then we can evolve the \perts\ using the transfer functions 
introduced in Section~\ref{transfncsec}.  The use of standard harmonic 
expansions at early times will, \eg, allow us to answer statistical 
questions about the \perts\ at late times, using the standard formalism 
of Gaussian random fields, just as we do on FRW backgrounds.

   To see how this works explicitly, consider first the comoving 
curvature \pert\ $\R$.  At the initial time $t_i$, when the fluctuations 
are well described by linear \perts\ from FRW, we can expand $\R$ in 
spherical harmonics as
\beq
\R(t_i,r,n^\mu) = \sqrt{\fr{2}{\pi}} \int dk\;k \sum_{\ell m} 
   \R_{\ell m}(k,t_i) j_\ell(kr) Y_{\ell m}(n^\mu),
\label{Rsphharm}
\eeq
where $j_\ell$ is a spherical Bessel function of the first kind and 
$k$ is a comoving wave number.  The curvature \pert\ $\R$ evaluated at 
time $t_i$, during matter domination, is related to the {\em primordial,} 
constant value $\Rpr$ via the standard matter transfer function $T(k)$:
\beq
\R_{\ell m}(k,t_i) = T(k)\Rpr_{\ell m}(k)
\label{RRpr}
\eeq
(see, \eg, \cite{ll00}).  The function $T(k)$ describes the suppression 
of power in $\R$ incurred while each mode is inside the Hubble radius 
during radiation domination.  Assuming Gaussian random primordial 
fluctuations, the statistics of $\Rpr$ are entirely encoded in the 
standard expression
\beq
\bra\Rpr_{\ell m}(k) \mathcal{R}^{\rm pr\ast}_{\ell' m'}(k')\ket = 
   2\pi^2\fr{\PR(k)}{k^3}\delta(k - k')\delta_{\ell\ell'}\delta_{mm'}.
\label{Rprstat}
\eeq
Here the angled brackets indicate averaging over the ensemble of 
realizations of the fluctuations.  For a scale invariant primordial 
spectrum the dimensionless power spectrum $\PR(k)$ is constant.

   Taking the energy density as an example, by combining 
\eqs(\ref{rholinIC}), (\ref{shearlinIC}), and (\ref{delrhotransfnc}) to 
(\ref{RRpr}), we can write\begin{widetext}
\beq
\fr{\del\rho(t,r,n^\mu)}{\rho(t,r)} =
   \fr{2}{5}\sqrt{\fr{2}{\pi}} \int dk\;k T(k) 
   \sum_{\ell m} \Rpr_{\ell m}(k) T_\rho(t,r,k,\ell) Y_{\ell m}(n^\mu),
\label{delrhofinal}
\eeq
where I have defined a new transfer function by
\beq
T_\rho(t,r,k,\ell) \equiv \ld(\fr{k}{a_iH_i}\rd)^2
   \ld[T_{\rho\rho}(t,r)j_\ell(kr)
     - T_{\rho\Sig}(t,r)\ld(j''_\ell(kr) + \oo{3}j_\ell(kr)\rd)\rd].
\label{newtransfnc}
\eeq
\end{widetext}
Here I have explicitly included the scale factor $a_i$ at time $t_i$, 
and $H_i \equiv \theta(t_i)/3$.  The notation $j''_\ell$ indicates the 
second derivative of $j_\ell$ with respect to its argument.  
Note that the details of a particular void 
profile enter these expressions only through the LTB transfer functions 
$T_{\rho\rho}(t,r)$ and $T_{\rho\Sig}(t,r)$.  I stress that the decomposition 
into wave numbers $k$ in these expressions has been done at the early time 
$t_i$, so the harmonic expansion is valid.

   We can now calculate any statistical aspect of the $\del\rho$ field, 
at any time.  For example, consider the angular power spectrum of 
density fluctuations on a sphere of some particular radius $r$ and at 
some particular time $t$, defined by
\beq
C_\ell^{\del\rho}(t,r)
    \equiv \ld\bra\int\fr{\del\rho(t,r,n^\mu)}{\rho(t,r)}
            Y_{\ell m}^\ast(n^\mu) d\Omega\rd\ket^2
\eeq
[presumably $r = r(t)$ if we are interested in observations on the past 
light cone].  Using \eqs(\ref{Rprstat}) and (\ref{delrhofinal}), and the 
orthonormality property of the spherical harmonics, we can evaluate 
this expression as
\beq
C_\ell^{\del\rho}(t,r) = \fr{16\pi}{25}
                         \int\fr{dk}{k}T^2(k)T^2_\rho(t,r,k,\ell)\PR(k).
\eeq
Using similar techniques we can calculate a large variety of observable 
quantities, which allows us to confront void models of acceleration 
with observations at the level of \perts\ \cite{voidsinprep}.

\section{Numerical evolution}
\label{numevsec}

   It is now a straightforward exercise to calculate any of the LTB transfer 
functions defined above.  First, we must specify an LTB background about 
which to perturb.  For numerical convenience I have used the function 
$K(r)$ defined in Sec.~\ref{exactsolnsec} to specify the intial 
profile, using the exact solution, \eqs(\ref{rhoexact}) to (\ref{Ysoln}).  
As discussed in the \hyperref[growdecayapp]{appendix}, I have set the 
bang time function $t_B(r)$ to zero to eliminate the decaying mode.  
The remaining radial gauge freedom was used to set $M(r) = r^3$, which 
is the value the corresponding function would take in FRW; although 
as discussed in Sec.~\ref{pertdefnsec} any choice of radial coordinate 
that agrees with the background at zeroth order would do.

   For the initial profile I chose
\beq
K(r) = \ld\{ \begin{array}{ll}
\displaystyle K_m\ld[\ld(\fr{r}{L}\rd)^5 - \fr{9}{5}\ld(\fr{r}{L}\rd)^4
      + \ld(\fr{r}{L}\rd)^2\rd], & \quad r \le L,\\
\displaystyle\oo{5}K_m\fr{L}{r}, & \quad r > L.
\end{array} \rd.
\eeq
This profile is characterized by two parameters, a width $L$ and an 
amplitude $K_m$.  For $r \gg L$, it approaches flat, matter-dominated 
FRW, \ie\ the Einstein-de~Sitter (EdS) universe.  I chose units such 
that the initial time was $t_i = 1$, and an amplitude which gave a spatial 
curvature of $\Rs/\theta^2 = -5.5\times10^{-4}$ at $t_i$, so that the 
initial profile is a small \pert\ from FRW.  The profile was chosen to 
roughly fit the luminosity distance-redshift relation for standard 
$\Lambda$CDM models, while providing a reasonable density parameter 
$\Om_m$ today.  However, evaluating the goodness of fit of the current 
LTB model to data is beyond the scope of this paper, and will be the 
subject of a future detailed study \cite{voidsinprep}.

\begin{figure}[ht]\begin{center}
\includegraphics[width=\columnwidth]{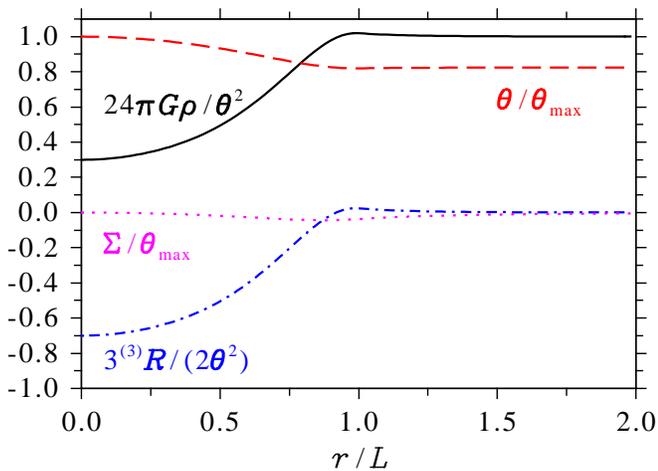}
\caption{Background LTB radial profile at the late time $t = 1.0\times10^5$.  
Energy density $\rho$, volume expansion rate $\theta$, shear $2$-scalar 
$\Sig$, and spatial curvature $\Rs$ are scaled as described in the text.  
This profile corresponds to an underdensity near the origin with effective 
density and curvature parameters $\Om_m = 0.3$ and $\Om_K = 0.7$.}
\label{bgndfig}
\end{center}\end{figure}

   In \fig\ref{bgndfig}, I display the background profiles for the energy 
density, expansion rate, $2$-scalar shear, and spatial curvature $\Rs$ 
calculated from the linearized Einstein energy constraint equation,
\beq
\fr{\theta^2}{3} = 8\pi G\rho - \half\Rs + \fr{3}{4}\Sig^2.
\label{enconstrexact}
\eeq
All quantities are displayed at the late time $t = 1.0\times10^5$, and were 
calculated using the exact solution, \eqs(\ref{rhoexact}) to (\ref{Ysoln}).  
In the plot, the expansion and shear $2$-scalar are normalized by the value 
of the expansion rate at the origin, $\theta_{\rm max}$.  The energy density 
and spatial curvature are normalized such that their difference should be 
unity, according to \eq(\ref{enconstrexact}) (the shear squared term is 
small compared to the density and curvature).

   Note that the normalized energy density and spatial curvature plotted 
in \fig\ref{bgndfig} correspond directly to the FRW definitions of the matter 
and (negative of the) curvature parameters, $\Om_m$ and $\Om_K$.  Therefore 
we can see that this LTB profile corresponds, at the late time plotted, to a 
significant underdensity extending from the centre to $r \simeq L$, with 
effective density and curvature parameters $\Om_m = 0.3$ and $\Om_K = 0.7$ 
at the centre.  Corresponding to this underdensity is a higher expansion 
rate than the asymptotic flat region.  The shear is small except at the 
boundary region $r \sim L$; it must vanish at the centre and in the 
asymptotic FRW region by symmetry.

\begin{figure}[ht]\begin{center}
\includegraphics[width=\columnwidth]{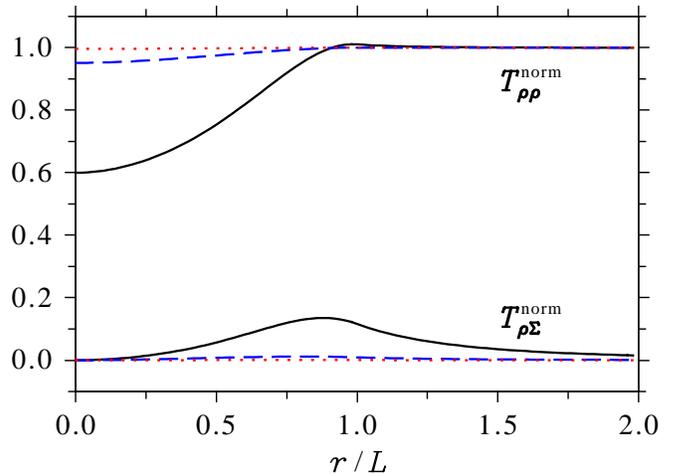}
\caption{Transfer functions $T_{\rho\rho}$ and $T_{\rho\Sig}$, normalized 
so that $T_{\rho\rho}^{\rm norm} = 1$ at all times for an 
Einstein-de~Sitter background.  The functions are presented for the times 
$t = 46.4$ (red, dotted line), $t = 2154$ (blue, dashed line), and 
$t = 1.0\times10^5$ 
(black, solid line).  Significant suppression of \perts\ occurs at late times 
within the void, and density \perts\ are also sourced by shear fluctuations 
at late times through $T_{\rho\Sig}$.}
\label{trnfncfig}
\end{center}\end{figure}

   Figure \ref{trnfncfig} shows the transfer functions $T_{\rho\rho}$ and 
$T_{\rho\Sig}$, which determine the evolution of the density \pert\ 
according to \eq(\ref{delrhotransfnc}).  They were calculated by 
numerically evolving the linearized set, \eqs(\ref{denconslin}) to 
(\ref{dEevolnlin}).  They have been normalized via
\beq
T_{\rho\rho}^{\rm norm}(t,r) \equiv \ld(\fr{t_i}{t}\rd)^{2/3}T_{\rho\rho}(t,r),
\eeq
and similarly for $T_{\rho\Sig}$.  This normalization was chosen so that, 
for the EdS case, $T_{\rho\rho}^{\rm norm}(t) = 1$ [and 
$T_{\rho\Sig}^{\rm norm}(t) = 0$] for all time.  The transfer functions 
are shown at the logarithmically-spaced times $t = 46.4$, $2154$, and 
$1.0\times10^5$.

   We can see from \fig\ref{trnfncfig} that at early times, and for 
$r \gg L$, the transfer functions approach their EdS values.  That is, 
the density \pert\ is evolving as expected when the background is (locally) 
nearly EdS.  However, we see a significant suppression in the growth of 
density \perts\ at late times within the void.  This suppression has 
an intuitive origin: near the centre of the void, the spacetime 
approximates the open FRW geometry, for which it is known that the growth 
of structure is suppressed compared with EdS (see, \eg, \cite{ll00}).  
It is possible to verify that the amount of suppression at the origin 
quantitatively matches the open FRW prediction \cite{voidsinprep}.

   The function $T_{\rho\Sig}$ plotted in \fig\ref{trnfncfig} illustrates 
that, at late times in the nonlinear background regime, shear fluctuations 
can source density fluctuations, via coupling through the background shear 
[recall \eqs(\ref{denconslin}) and (\ref{dRaychaudlin})].  The importance 
of this effect is not clear from this figure, however, since the two 
transfer functions $T_{\rho\rho}$ and $T_{\rho\Sig}$ multiply different 
combinations of Bessel functions in \eq(\ref{newtransfnc}).

\begin{figure}[ht]\begin{center}
\includegraphics[width=\columnwidth]{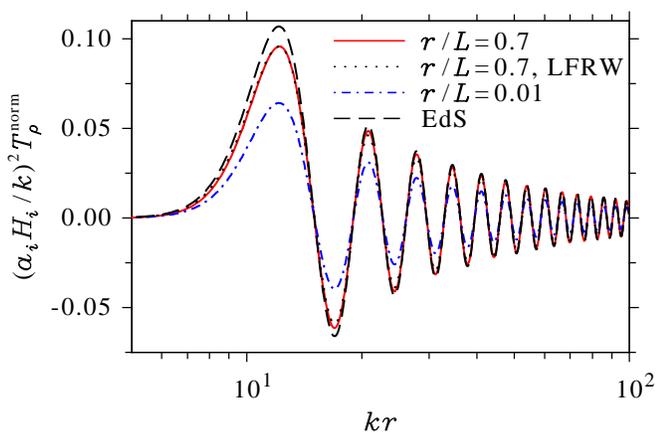}
\caption{Normalized transfer function $T_\rho^{\rm norm}(t,r,k,\ell)$ 
for $\ell = 10$.  The curves are shown for a point within the periphery 
of the LTB void, at $r/L = 0.7$, as well as near the void centre at 
$r/L = 0.01$.  Also shown is the local FRW approximation (LFRW) for 
$r/L = 0.7$, together with the Einstein-de~Sitter case (EdS).}
\label{matangtffig}
\end{center}\end{figure}

   Finally, \fig\ref{matangtffig} illustrates the transfer function 
$T_\rho(t,r,k,\ell)$, which was defined in \eq(\ref{newtransfnc}), again 
normalized according to
\beq
T_\rho^{\rm norm}(t,r,k,\ell)
   \equiv \ld(\fr{t_i}{t}\rd)^{2/3}T_\rho(t,r,k,\ell),
\eeq
for the case $\ell = 10$ (the general features are very similar for other 
values of $\ell$).  With this normalization, we have $T_\rho(t,r,k,\ell) = 
(k/a_iH_i)^2 j_\ell(kr)$ for all $t$ and $r$ for the special case of the 
EdS universe, and this case is plotted in \fig\ref{matangtffig}.  Also 
plotted are the curves for the LTB model defined above, and 
evaluated within the periphery of the void, at $r/L = 0.7$, and near the 
void centre, at $r/L = 0.01$ (cf. \fig\ref{trnfncfig}).  For the LTB cases, 
the transfer function is evaluated on the past light cone.  
Figure~\ref{matangtffig} illustrates again the suppression of matter 
fluctuations near the origin with respect to the EdS case.

   To illustrate the effect of the background shear $\Sig$ and tidal field 
$\E$, which are non-negligible only within the peripheral region of the void 
(recall \fig\ref{bgndfig}), \fig\ref{matangtffig} also presents the transfer 
function $T_\rho(t,r,k,\ell)$ evaluated at $r/L = 0.7$, but calculated 
by ignoring the background shear and tidal field in the linearized evolution 
\eqs(\ref{denconslin}) to (\ref{dEevolnlin}).  This approximation can 
be called the ``local FRW approximation,'' since setting $\Sig = \E = 0$ 
results in the evolution of the local quantities (background and \perts) 
along each worldline that is identical to that of some FRW universe.  We see 
that the local FRW approximation is very good, even in the void periphery, 
where the background shear and tidal field are largest.  (The local FRW 
approximation transfer function for $r/L = 0.01$ is indistinguishable from 
the corresponding curve plotted in \fig\ref{matangtffig}.)  Note that in 
the local FRW approximation we have $T_{\rho\Sig} = 0$, so 
\fig\ref{matangtffig} shows that modifications to the matter power 
spectrum due to shear fluctuations (as encoded in the transfer function 
$T_{\rho\Sig}$) are subdominant to the main suppression which is captured 
well by the local FRW approximation.

   Recall that the tensor modes, which we have ignored in all of these 
calculations, couple to $\del\rho$ only through the shear [see 
\eqs(\ref{enconslin}) to (\ref{Eevolnlin})].  Since the magnitude of the 
background shear does not depend on the amplitude of linear tensor \perts, 
the weak effect of the shear that we have observed (\ie\ the excellent 
accuracy of the local FRW approximation) means that it is very 
unlikely that the tensors would have a significant effect on the density 
fluctuations had the tensors been included.  It should be stressed that 
this result will depend on the background profile:  a background more 
strongly nonlinear, and hence with larger shear, than the one studied 
here will couple both scalar shear fluctuations and tensor modes more 
strongly to the density \perts.  Thus the importance of the background 
shear should be evaluated whenever evolving scalar \perts\ on an LTB 
background.  As well, the effect of the background shear may be larger 
on quantities other than the density \perts.

\section{Discussion}

   It is expected that the formalism developed here could be applied 
to constraining void models for acceleration with a wide variety of 
structure-related data.  Such a program could perhaps be dubbed ``testing 
homogeneity with inhomogeneities.''  Examples include the shape (and 
amplitude) of the local matter power spectrum, a rigorous treatment 
of baryon acoustic oscillations, and the effects of gravitational lensing 
by linear structures.  These applications are currently under study 
\cite{voidsinprep}.  Apart from void models, it is expected that the 
present approach could find use elsewhere, in particular in examining 
the early stages of the standard structure formation process.

   A particularly interesting application would be a full calculation 
of the effects of a void inhomogeneity on the CMB.  The effect of 
varying the angular diameter distance to the last scattering surface in 
standard FRW cosmologies is a change in the angular peak scale (indeed, 
this is the dominant effect on the spectrum from varying the {\em time} 
of observation, keeping everything else fixed \cite{zms07}), which can 
already be calculated at background level in LTB models \cite{aag06}.  The 
free-streamed primary CMB anisotropy power spectrum is expected to be 
very robust to the presence of isotropic inhomogeneity.  However, 
recall from \fig\ref{trnfncfig} that the presence of a large nonlinear 
void results in a {\em suppression} of the power of scalar fluctuations 
inside the void.  Such a suppression exists at late times in standard 
$\Lambda$CDM 
models (as well as in open models), where it leads to the integrated 
Sachs-Wolfe (ISW) effect.  An analogous effect should occur in 
void models, which will provide another handle on constraining these 
models.  The ISW effect modifies the largest angular scales, and so 
its presence is difficult to distinguish directly due to cosmic variance.  
However, correlations with large-scale structure have been measured 
with high confidence \cite{hhpsb08}, and this is a signature that the 
present formalism could be used to calculate for void models.  Even if 
the overall {\em amplitude} of the ISW-structure correlations turns 
out to be insufficiently constraining, it would be very surprizing if 
the redshift-dependence of the effect would be similar between void 
and $\Lambda$CDM models.

   It is important to stress again that the present approach is grounded 
within the well-developed theory of linear \perts\ on FRW backgrounds, 
and many of the familiar standard techniques from that field can be 
applied here.  The main difference is the presence of a small number 
of new, simple to calculate linear transfer functions.  In practical 
terms, these transfer functions only need to be calculated once for 
a particular void profile, and can be expressed as a function of a 
single variable if results are required on the past light cone, as is 
normally the case.  This connection with standard FRW \pert\ theory 
was made possible by the crucial assumption that voids contain no 
decaying modes, so that they are consistent with the early homogeneity 
predicted by models of inflation.

   On the theoretical side, future work could include the calculation 
of the remaining $2$-vector and $2$-tensor shear and electric Weyl 
degrees of freedom, which, in the absence of coupling to $H_\mn$, also 
evolve according to ODEs, as \eqs(\ref{Sigevoln}) and (\ref{Eevoln}) 
show.  Apart from possible observational consequences of these components, 
this would enable a consistency check on the accuracy of ignoring the 
tensor coupling by monitoring the irrotational constraint equation 
\cite{tcm08}
\beq
H_\mn = {\rm curl}\sig_\mn.
\eeq

   However, it is important to stress that, for the void model studied 
here, the effect of the background shear on the density fluctuations was 
found to be weak.  Thus any tensors produced are not expected to have 
a significant effect on the density \perts.  The dominant effect, namely 
a significant suppression of power within the void, was found to be 
captured well by the local FRW approximation, which ignores the 
background shear and tidal field.  Crucially, this dominant suppression 
near the origin is independent of the details of the void profile away 
from the origin, since the background shear must vanish at the origin by 
isotropy, so that the local FRW approximation is exact there.  Similarly, 
the degree of suppression near the centre will not be affected by tensors, 
which require background shear to couple to density fluctuations.

   To conclude, proponents for void scenarios argue that they are 
conservative in that they require no new, mysterious dark energy 
component (although, of course, they leave the solution of the 
cosmological constant problem to future work!), and that they may 
naturally explain the ``coincidence problem,'' since structure started 
to go nonlinear only recently.  Ultimately, though, we will need to 
resort to the widest range of observations possible to determine if the 
apparently natural philosophical stance of the Copernican principle 
actually reflects our place in the Universe.

\begin{acknowledgments}
This research was supported by the Natural Sciences and Engineering Research 
Council of Canada.  I thank Adam Moss and Douglas Scott for useful 
discussions.
\end{acknowledgments}

\appendix*
\section{Growing and decaying modes}
\label{growdecayapp}
\subsection{Linearized FRW solution}

   We saw in Sec.~\ref{exactsolnsec} that the general spherically 
symmetric dust (LTB) spacetime is specified by two free radial functions.  
To help elucidate the physical significance of these two free functions, 
consider an LTB spacetime which is very ``close'' to FRW, in the sense 
that it can be treated as a linear \pert\ from FRW.  It is well known 
from the theory of cosmological \perts\ (see, \eg, \cite{mfb92}) that a 
general linear dust \pert\ on FRW is described by two modes, one growing 
and one decaying.  [This can be seen easily from the exact 
\eqs(\ref{encons}) and (\ref{Raychaud}):  For an FRW background, the 
shear term $\sig_\mn\sig^\mn$ is second order, and hence the linearized 
\eqs(\ref{encons}) and (\ref{Raychaud}) decouple from the remaining 
evolution equations and form a second order system with two independent 
solutions.]  Therefore we can conclude that the two free functions of an 
LTB spacetime correspond to growing and decaying modes in the linear 
regime about FRW.  This point was made some time ago in Ref.~\cite{silk77} 
(see also \cite{hk06}), but it will be useful to derive the result here 
in a more explicit manner.

   The importance of this result is that the presence of a significant 
decaying mode at {\em late} times will result in extreme inhomogeneity at 
{\em early} times, contradicting the standard inflationary scenario, 
according to which the Universe is expected to be extremely \homo\ at the 
end of inflation.  Hence we will need to know how to set the decaying 
mode to zero in setting up a spherical LTB profile.  This reduces the 
freedom available in choosing a profile, and hence should make it more 
difficult to fit all of the data.

   The evolution of the {\em comoving gauge} dust energy density \pert\ 
$\del\rho$ in a pressureless FRW background is given by
\beq
\fr{\del\rho}{\rho} = \ld(\fr{t}{t_0}\rd)^{2/3}D_1(x) + \fr{t_0}{t}D_2(x),
\label{delrhoFRW}
\eeq
where $D_1(x)$ and $D_2(x)$ are the amplitudes of the growing and 
decaying modes, respectively, and $t_0$ is an arbitrary reference time.  
[It is straightforward to show that this expression is the solution to 
the linearized \eqs(\ref{encons}) and (\ref{Raychaud}).]  
Similarly, it can be shown that the comoving gauge expansion \pert\ 
$\del\theta$ \footnote{Precisely, the \pert\ $\del\theta$ is defined 
as the \pert\ in the expansion of the comoving worldlines, evaluated 
on the comoving-orthogonal slices.} is given by a growing and decaying mode,
\beq
\fr{\del\theta}{\theta} = -\oo{3}\ld(\fr{t}{t_0}\rd)^{2/3}D_1(x)
                        + \half\fr{t_0}{t}D_2(x).
\label{delthetaFRW}
\eeq
Using these two expressions and the linearized energy constraint equation, 
we can show that the $3$-Ricci curvature of the comoving slices, $\Rs$, 
evolves according to
\beq
-\fr{3}{2}\fr{\Rs}{\theta^2} = -\fr{5}{3}\ld(\fr{t}{t_0}\rd)^{2/3}D_1(x).
\label{delRsFRW}
\eeq
Note that the curvature \pert\ consists of {\em just the growing mode.}  
The normalization of the $3$-curvature \pert\ in \eq(\ref{delRsFRW}) has 
been chosen to agree with the conventional FRW curvature parameter,
\beq
\Om_K = -\fr{3}{2}\fr{\Rs}{\theta^2}
\eeq
in the \homo\ limit.

\subsection{Vanishing growing mode}

   I showed above that a comoving $3$-curvature \pert\ corresponds to a 
pure {\em growing} mode in linear theory on an FRW background.  Therefore 
we can isolate the {\em decaying} mode by considering solutions with 
vanishing spatial Ricci curvature $\Rs$.  For an LTB spacetime which 
is a small \pert\ from FRW, the exact energy conservation and Raychaudhuri 
equations, (\ref{encons}) and (\ref{Raychaud}), become, to first order,
\beq
\dot\rho = -\theta\rho,
\label{enconsgd}
\eeq
\beq
\dot\theta = -\fr{\theta^2}{3} - 4\pi G\rho,
\eeq
since the shear $\sig_\mn$ is a first-order quantity.  Under the same 
approximation, the Einstein energy constraint equation becomes
\beq
\fr{\theta^2}{3} = 8\pi G\rho - \half\Rs.
\label{enconstrgd}
\eeq
For the case $\Rs = 0$, it is simple to verify that the solution to 
these equations is
\beq
\rho(r,t) = \oo{6\pi G[t - t_B(r)]^2},
\label{rhodecay}
\eeq
\beq
\theta(r,t) = \fr{2}{t - t_B(r)}.
\label{thetadecay}
\eeq
Here the function $t_B(r)$ has an arbitrary radial profile, but in order 
that the shear (and electric Weyl tensor) be small, the {\em size} of 
radial fluctuations in $t_B(r)$ must not be too large.

   To quantify this, consider a \homo\ background solution
\beq
\rho^{(0)}(t) = \oo{6\pi G\ld(t - t^{(0)}_B\rd)^2},
\eeq
\beq
\theta^{(0)}(t) = \fr{2}{t - t^{(0)}_B},
\eeq
near the perturbed solution (\ref{rhodecay}) and (\ref{thetadecay}), where 
$t^{(0)}_B$ is a constant.  Define the energy density \pert\ by
\beq
\Delta\rho(r,t) \equiv \rho(r,t) - \rho^{(0)}(t).
\label{delrhoLTB}
\eeq
Then, expanding in powers of $\Delta t_B/t$, where $\Delta t_B(r) 
\equiv t_B(r) - t^{(0)}_B$, we have
\beq
\fr{\Delta\rho}{\rho^{(0)}} = 2\fr{\Delta t_B}{t_0}\fr{t_0}{t}
                            + \O\ld(\fr{\Delta t_B}{t}\rd)^2,
\label{DelrhodecayLTB}
\eeq
with $t_0$ an arbitrary 
reference time.  Comparing this expression with \eq(\ref{delrhoFRW}), we 
see that we have indeed isolated the decaying mode, and its amplitude 
can be read off of \eq(\ref{DelrhodecayLTB}) as
\beq
D_2(r) = 2\fr{\Delta t_B(r)}{t_0}.
\label{D2LTB}
\eeq

   A similar calculation gives
\beq
\fr{\Delta\theta}{\theta^{(0)}} = \fr{\Delta t_B}{t_0}\fr{t_0}{t}
                                + \O\ld(\fr{\Delta t_B}{t}\rd)^2,
\eeq
which, with \eq(\ref{delthetaFRW}), again gives \eq(\ref{D2LTB}).  Therefore 
it is {\em fluctuations} of the bang time $t_B(r)$ that give the decaying 
mode, and we can set the decaying mode to zero by setting $t_B(r) = \const$.

   It is very important to point out that the comparison between the 
ordinary FRW \perts\ $\del\rho$ and $\del\theta$ and the \perts\ 
$\Delta\rho$ and $\Delta\theta$ is valid, since the FRW \perts\ were 
specified in comoving gauge, and the LTB solutions \eq(\ref{rhodecay}) 
and (\ref{thetadecay}) are specified with respect to the comoving congruence.

   Finally, note that at times early enough that $t \simeq \Delta t_B(r)$, 
the linear approximation breaks down.  This is expected, since the decaying 
mode diverges at early times.  This result is intuitively clear, as we expect 
differences in ``bang time'' to become all-important at early times, but 
to be inconsequential at sufficiently late times.

\subsection{Vanishing decaying mode}

   Now that we have isolated the LTB decaying mode, we can examine the 
growing mode, by looking at solutions with $t_B(r) = 0$, but with 
nonvanishing curvature of the comoving slices.  For definiteness I will 
consider the open case, $\Rs < 0$; the other cases follow similarly.

   In this case, \eqs(\ref{enconsgd}) to (\ref{enconstrgd}) have the solution
\beq
\rho(r,t) = \rho_0(r)\ld(\fr{a_0}{a}\rd)^3,
\label{rhogrow}
\eeq
\beq
\theta(r,t) = 3\fr{\dot a}{a},
\eeq
where the quantity $a = a(r,t)$ is given parametrically by
\begin{subequations}
\label{paramgd}
\beq
\fr{a}{a_0} = \fr{1 - \Om_{K,0}}{2\Om_{K,0}}(\cosh\eta - 1),
\eeq
\beq
\theta_0(r)t = \fr{3}{2}\fr{1 - \Om_{K,0}}{\Om_{K,0}^{3/2}}(\sinh\eta - \eta).
\label{paramtgd}
\eeq
\end{subequations}
Quantities with subscript $0$ are evaluated at reference time $t = t_0$, 
and are related by
\beq
(1 - \Om_{K,0})\theta_0^2 = 24\pi G\rho_0,
\eeq
and, by analogy with the \homo\ FRW case, I define
\beq
\Om_{K,0}(r) \equiv -\fr{3}{2}\fr{\Rs_0}{\theta_0^2}.
\label{OmKdef}
\eeq
Therefore the solution is determined by one free radial function, which 
describes the spatial Ricci curvature on some reference comoving slice.

   Again, consider a spatially flat \homo\ background solution
\beq
\rho^{(0)}(t) = \oo{6\pi Gt^2},
\eeq
and define the density \pert\ $\Delta\rho$ by \eq(\ref{delrhoLTB}).  If 
we are to evaluate this \pert\ at time $t_0$, then for the linear 
approximation to be valid we require $\Om_{K,0}(r) \ll 1$, \ie\ the 
\pert\ must correspond to a small curvature fluctuation.  Then 
\eq(\ref{paramtgd}) tells us that $\eta \ll 1$ (this is equivalent to 
the ``small $u$'' approximation in \cite{bn07}).  We can now solve the set 
\eq(\ref{paramgd}) for $a/a_0$ perturbatively in $\eta$.  Keeping the 
next-to-leading order terms in the expansions for the hyperbolic 
functions, I find
\beq
\fr{a}{a_0} = \ld(6\pi G\rho_0t^2\rd)^{1/3}
            + \oo{80}\fr{1 - \Om_{K,0}}{\Om_{K,0}}\eta^4 + \O(\eta^6).
\eeq
Combining this with the lowest-order expression
\beq
\theta_0t = \fr{1 - \Om_{K,0}}{4\Om_{K,0}^{3/2}}\eta^3 + \O(\eta^5)
\eeq
gives
\beq
\fr{a}{a_0} = \ld(6\pi G\rho_0t^2\rd)^{1/3} \ld[1 +
   \oo{5}\fr{\Om_{K,0}}{(1 - \Om_{K,0})^{2/3}}\ld(\fr{t}{t_0}\rd)^{2/3}\rd].
\eeq
Combining this expression with \eq(\ref{rhogrow}) gives
\beq
\rho(r,t) = \oo{6\pi Gt^2} \ld[1 - \fr{3}{5}
   \fr{\Om_{K,0}}{(1 - \Om_{K,0})^{2/3}}\ld(\fr{t}{t_0}\rd)^{2/3}\rd].
\eeq
Therefore the density \pert\ is finally determined to be
\beq
\fr{\Delta\rho}{\rho^{(0)}} = -\fr{3}{5}\Om_{K,0}\ld(\fr{t}{t_0}\rd)^{2/3}
\label{DelrhogrowLTB}
\eeq
at lowest order.  Comparing with \eq(\ref{delrhoFRW}), we see that, as 
expected, \eq(\ref{DelrhogrowLTB}) consists of a pure growing mode.  As 
with the decaying mode case, we can read off the growing mode amplitude as
\beq
D_1(r) = -\fr{3}{5}\Om_{K,0}.
\label{D1LTB}
\eeq
With \eq(\ref{OmKdef}), we recover \eq(\ref{delRsFRW}), \ie\ the growing 
mode is given by the spatial curvature, as expected.

   A similar calculation gives
\beq
\fr{\Delta\theta}{\theta^{(0)}} = \oo{5}\Om_{K,0}\ld(\fr{t}{t_0}\rd)^{2/3},
\eeq
which, with \eq(\ref{D1LTB}), is again consistent with the FRW expression 
for the growing expansion \pert\ mode, \eq(\ref{delthetaFRW}).  Therefore, 
the growing mode for LTB spacetimes near FRW is indeed given by the 
spatial curvature of the comoving slices.

\bibliography{bib}

\begin{thebibliography}{39}
\expandafter\ifx\csname natexlab\endcsname\relax\def\natexlab#1{#1}\fi
\expandafter\ifx\csname bibnamefont\endcsname\relax
  \def\bibnamefont#1{#1}\fi
\expandafter\ifx\csname bibfnamefont\endcsname\relax
  \def\bibfnamefont#1{#1}\fi
\expandafter\ifx\csname citenamefont\endcsname\relax
  \def\citenamefont#1{#1}\fi
\expandafter\ifx\csname url\endcsname\relax
  \def\url#1{\texttt{#1}}\fi
\expandafter\ifx\csname urlprefix\endcsname\relax\def\urlprefix{URL }\fi
\providecommand{\bibinfo}[2]{#2}
\providecommand{\eprint}[2][]{\url{#2}}

\bibitem[{\citenamefont{Hogg et~al.}(2005)}]{hogg04}
\bibinfo{author}{\bibfnamefont{D.~W.} \bibnamefont{Hogg}} \bibnamefont{et~al.},
  \bibinfo{journal}{Astrophys. J.} \textbf{\bibinfo{volume}{624}},
  \bibinfo{pages}{54} (\bibinfo{year}{2005}), arXiv:astro-ph/0411197.

\bibitem[{\citenamefont{Lahav}(2002)}]{lahav02}
\bibinfo{author}{\bibfnamefont{O.}~\bibnamefont{Lahav}},
  \bibinfo{journal}{Class. Quant. Grav.} \textbf{\bibinfo{volume}{19}},
  \bibinfo{pages}{3517} (\bibinfo{year}{2002}), arXiv:astro-ph/0112524.

\bibitem[{\citenamefont{Goodman}(1995)}]{goodman95}
\bibinfo{author}{\bibfnamefont{J.}~\bibnamefont{Goodman}},
  \bibinfo{journal}{Phys. Rev.} \textbf{\bibinfo{volume}{D52}},
  \bibinfo{pages}{1821} (\bibinfo{year}{1995}), arXiv:astro-ph/9506068.

\bibitem[{\citenamefont{Caldwell and Stebbins}(2007)}]{cs07}
\bibinfo{author}{\bibfnamefont{R.~R.} \bibnamefont{Caldwell}} \bibnamefont{and}
  \bibinfo{author}{\bibfnamefont{A.}~\bibnamefont{Stebbins}}
  (\bibinfo{year}{2007}), arXiv:0711.3459 [astro-ph].

\bibitem[{\citenamefont{Clarkson et~al.}(2007)\citenamefont{Clarkson, Bassett,
  and Lu}}]{cbl07}
\bibinfo{author}{\bibfnamefont{C.}~\bibnamefont{Clarkson}},
  \bibinfo{author}{\bibfnamefont{B.~A.} \bibnamefont{Bassett}},
  \bibnamefont{and} \bibinfo{author}{\bibfnamefont{T.~H.-C.} \bibnamefont{Lu}}
  (\bibinfo{year}{2007}), arXiv:0712.3457 [astro-ph].

\bibitem[{\citenamefont{Uzan et~al.}(2008)\citenamefont{Uzan, Clarkson, and
  Ellis}}]{uce08}
\bibinfo{author}{\bibfnamefont{J.-P.} \bibnamefont{Uzan}},
  \bibinfo{author}{\bibfnamefont{C.}~\bibnamefont{Clarkson}}, \bibnamefont{and}
  \bibinfo{author}{\bibfnamefont{G.~F.~R.} \bibnamefont{Ellis}},
  \bibinfo{journal}{Phys. Rev. Lett.} \textbf{\bibinfo{volume}{100}},
  \bibinfo{pages}{191303} (\bibinfo{year}{2008}), arXiv:0801.0068 [astro-ph].

\bibitem[{\citenamefont{Celerier}(2000)}]{celerier99}
\bibinfo{author}{\bibfnamefont{M.-N.} \bibnamefont{Celerier}},
  \bibinfo{journal}{Astron. Astrophys.} \textbf{\bibinfo{volume}{353}},
  \bibinfo{pages}{63} (\bibinfo{year}{2000}), arXiv:astro-ph/9907206.

\bibitem[{\citenamefont{{Lema{\^i}tre}}(1933)}]{lemaitre33}
\bibinfo{author}{\bibfnamefont{G.}~\bibnamefont{{Lema{\^i}tre}}},
  \bibinfo{journal}{Annales de la Societe Scietifique de Bruxelles}
  \textbf{\bibinfo{volume}{53}}, \bibinfo{pages}{51} (\bibinfo{year}{1933}).

\bibitem[{\citenamefont{Tolman}(1934)}]{tolman34}
\bibinfo{author}{\bibfnamefont{R.~C.} \bibnamefont{Tolman}},
  \bibinfo{journal}{Proc. Nat. Acad. Sci.} \textbf{\bibinfo{volume}{20}},
  \bibinfo{pages}{169} (\bibinfo{year}{1934}).

\bibitem[{\citenamefont{Bondi}(1947)}]{bondi47}
\bibinfo{author}{\bibfnamefont{H.}~\bibnamefont{Bondi}}, \bibinfo{journal}{Mon.
  Not. Roy. Astron. Soc.} \textbf{\bibinfo{volume}{107}}, \bibinfo{pages}{410}
  (\bibinfo{year}{1947}).

\bibitem[{\citenamefont{Alnes et~al.}(2006)\citenamefont{Alnes, Amarzguioui,
  and Gron}}]{aag06}
\bibinfo{author}{\bibfnamefont{H.}~\bibnamefont{Alnes}},
  \bibinfo{author}{\bibfnamefont{M.}~\bibnamefont{Amarzguioui}},
  \bibnamefont{and} \bibinfo{author}{\bibfnamefont{O.}~\bibnamefont{Gron}},
  \bibinfo{journal}{Phys. Rev.} \textbf{\bibinfo{volume}{D73}},
  \bibinfo{pages}{083519} (\bibinfo{year}{2006}), arXiv:astro-ph/0512006.

\bibitem[{\citenamefont{Enqvist and Mattsson}(2007)}]{em07}
\bibinfo{author}{\bibfnamefont{K.}~\bibnamefont{Enqvist}} \bibnamefont{and}
  \bibinfo{author}{\bibfnamefont{T.}~\bibnamefont{Mattsson}},
  \bibinfo{journal}{JCAP} \textbf{\bibinfo{volume}{0702}}, \bibinfo{pages}{019}
  (\bibinfo{year}{2007}), arXiv:astro-ph/0609120.

\bibitem[{\citenamefont{Alexander et~al.}(2007)\citenamefont{Alexander, Biswas,
  Notari, and Vaid}}]{abnv07}
\bibinfo{author}{\bibfnamefont{S.}~\bibnamefont{Alexander}},
  \bibinfo{author}{\bibfnamefont{T.}~\bibnamefont{Biswas}},
  \bibinfo{author}{\bibfnamefont{A.}~\bibnamefont{Notari}}, \bibnamefont{and}
  \bibinfo{author}{\bibfnamefont{D.}~\bibnamefont{Vaid}}
  (\bibinfo{year}{2007}), arXiv:0712.0370 [astro-ph].

\bibitem[{\citenamefont{Garcia-Bellido and Haugboelle}(2008)}]{gbh08}
\bibinfo{author}{\bibfnamefont{J.}~\bibnamefont{Garcia-Bellido}}
  \bibnamefont{and}
  \bibinfo{author}{\bibfnamefont{T.}~\bibnamefont{Haugboelle}},
  \bibinfo{journal}{JCAP} \textbf{\bibinfo{volume}{0804}}, \bibinfo{pages}{003}
  (\bibinfo{year}{2008}), arXiv:0802.1523 [astro-ph].

\bibitem[{\citenamefont{Enqvist}(2008)}]{enqvist08}
\bibinfo{author}{\bibfnamefont{K.}~\bibnamefont{Enqvist}},
  \bibinfo{journal}{Gen. Rel. Grav.} \textbf{\bibinfo{volume}{40}},
  \bibinfo{pages}{451} (\bibinfo{year}{2008}), arXiv:0709.2044 [astro-ph].

\bibitem[{\citenamefont{Greenberg}(1970)}]{greenberg70}
\bibinfo{author}{\bibfnamefont{P.~J.} \bibnamefont{Greenberg}},
  \bibinfo{journal}{J. Math. Anal. Appl.} \textbf{\bibinfo{volume}{30}},
  \bibinfo{pages}{128} (\bibinfo{year}{1970}).

\bibitem[{\citenamefont{Clarkson and Barrett}(2003)}]{cb03}
\bibinfo{author}{\bibfnamefont{C.~A.} \bibnamefont{Clarkson}} \bibnamefont{and}
  \bibinfo{author}{\bibfnamefont{R.~K.} \bibnamefont{Barrett}},
  \bibinfo{journal}{Class. Quant. Grav.} \textbf{\bibinfo{volume}{20}},
  \bibinfo{pages}{3855} (\bibinfo{year}{2003}), arXiv:gr-qc/0209051.

\bibitem[{\citenamefont{Clarkson}(2007)}]{clarkson07}
\bibinfo{author}{\bibfnamefont{C.}~\bibnamefont{Clarkson}},
  \bibinfo{journal}{Phys. Rev.} \textbf{\bibinfo{volume}{D76}},
  \bibinfo{pages}{104034} (\bibinfo{year}{2007}), arXiv:0708.1398 [gr-qc].

\bibitem[{\citenamefont{Tsagas et~al.}(2008)\citenamefont{Tsagas, Challinor,
  and Maartens}}]{tcm08}
\bibinfo{author}{\bibfnamefont{C.~G.} \bibnamefont{Tsagas}},
  \bibinfo{author}{\bibfnamefont{A.}~\bibnamefont{Challinor}},
  \bibnamefont{and} \bibinfo{author}{\bibfnamefont{R.}~\bibnamefont{Maartens}},
  \bibinfo{journal}{Phys. Rept. (to be published)}  (\bibinfo{year}{2008}),
  arXiv:0705.4397 [astro-ph].

\bibitem[{\citenamefont{Linde et~al.}(1995)\citenamefont{Linde, Linde, and
  Mezhlumian}}]{llm95}
\bibinfo{author}{\bibfnamefont{A.~D.} \bibnamefont{Linde}},
  \bibinfo{author}{\bibfnamefont{D.~A.} \bibnamefont{Linde}}, \bibnamefont{and}
  \bibinfo{author}{\bibfnamefont{A.}~\bibnamefont{Mezhlumian}},
  \bibinfo{journal}{Phys. Lett.} \textbf{\bibinfo{volume}{B345}},
  \bibinfo{pages}{203} (\bibinfo{year}{1995}), arXiv:hep-th/9411111.

\bibitem[{\citenamefont{Zibin and Moss}(2008)}]{voidsinprep}
\bibinfo{author}{\bibfnamefont{J.~P.} \bibnamefont{Zibin}} \bibnamefont{and}
  \bibinfo{author}{\bibfnamefont{A.}~\bibnamefont{Moss}}
  (\bibinfo{year}{2008}), \bibinfo{note}{in preparation}.

\bibitem[{\citenamefont{Wald}(1984)}]{wald84}
\bibinfo{author}{\bibfnamefont{R.~M.} \bibnamefont{Wald}},
  \emph{\bibinfo{title}{General Relativity}} (\bibinfo{publisher}{University of
  Chicago Press}, \bibinfo{address}{Chicago}, \bibinfo{year}{1984}).

\bibitem[{\citenamefont{van Elst and Ellis}(1996)}]{vee96}
\bibinfo{author}{\bibfnamefont{H.}~\bibnamefont{van Elst}} \bibnamefont{and}
  \bibinfo{author}{\bibfnamefont{G.~F.~R.} \bibnamefont{Ellis}},
  \bibinfo{journal}{Class. Quant. Grav.} \textbf{\bibinfo{volume}{13}},
  \bibinfo{pages}{1099} (\bibinfo{year}{1996}), arXiv:gr-qc/9510044.

\bibitem[{\citenamefont{Sussman and Garcia~Trujillo}(2002)}]{st02}
\bibinfo{author}{\bibfnamefont{R.~A.} \bibnamefont{Sussman}} \bibnamefont{and}
  \bibinfo{author}{\bibfnamefont{L.}~\bibnamefont{Garcia~Trujillo}},
  \bibinfo{journal}{Class. Quant. Grav.} \textbf{\bibinfo{volume}{19}},
  \bibinfo{pages}{2897} (\bibinfo{year}{2002}), arXiv:gr-qc/0105081.

\bibitem[{\citenamefont{Carbone et~al.}(2006)\citenamefont{Carbone,
  Baccigalupi, and Matarrese}}]{cbm07}
\bibinfo{author}{\bibfnamefont{C.}~\bibnamefont{Carbone}},
  \bibinfo{author}{\bibfnamefont{C.}~\bibnamefont{Baccigalupi}},
  \bibnamefont{and}
  \bibinfo{author}{\bibfnamefont{S.}~\bibnamefont{Matarrese}},
  \bibinfo{journal}{Phys. Rev.} \textbf{\bibinfo{volume}{D73}},
  \bibinfo{pages}{063503} (\bibinfo{year}{2006}), arXiv:astro-ph/0509680.

\bibitem[{\citenamefont{Mutoh et~al.}(1997)\citenamefont{Mutoh, Hirai, and
  Maeda}}]{mhm97}
\bibinfo{author}{\bibfnamefont{H.}~\bibnamefont{Mutoh}},
  \bibinfo{author}{\bibfnamefont{T.}~\bibnamefont{Hirai}}, \bibnamefont{and}
  \bibinfo{author}{\bibfnamefont{K.-i.} \bibnamefont{Maeda}},
  \bibinfo{journal}{Phys. Rev.} \textbf{\bibinfo{volume}{D55}},
  \bibinfo{pages}{3276} (\bibinfo{year}{1997}), arXiv:astro-ph/9608183.

\bibitem[{\citenamefont{Matarrese et~al.}(1993)\citenamefont{Matarrese,
  Pantano, and Saez}}]{mps93}
\bibinfo{author}{\bibfnamefont{S.}~\bibnamefont{Matarrese}},
  \bibinfo{author}{\bibfnamefont{O.}~\bibnamefont{Pantano}}, \bibnamefont{and}
  \bibinfo{author}{\bibfnamefont{D.}~\bibnamefont{Saez}},
  \bibinfo{journal}{Phys. Rev.} \textbf{\bibinfo{volume}{D47}},
  \bibinfo{pages}{1311} (\bibinfo{year}{1993}).

\bibitem[{\citenamefont{van Elst et~al.}(1997)\citenamefont{van Elst, Uggla,
  Lesame, Ellis, and Maartens}}]{veulem97}
\bibinfo{author}{\bibfnamefont{H.}~\bibnamefont{van Elst}},
  \bibinfo{author}{\bibfnamefont{C.}~\bibnamefont{Uggla}},
  \bibinfo{author}{\bibfnamefont{W.~M.} \bibnamefont{Lesame}},
  \bibinfo{author}{\bibfnamefont{G.~F.~R.} \bibnamefont{Ellis}},
  \bibnamefont{and} \bibinfo{author}{\bibfnamefont{R.}~\bibnamefont{Maartens}},
  \bibinfo{journal}{Class. Quant. Grav.} \textbf{\bibinfo{volume}{14}},
  \bibinfo{pages}{1151} (\bibinfo{year}{1997}), arXiv:gr-qc/9611002.

\bibitem[{\citenamefont{{Bardeen}}(1988)}]{bardeen88}
\bibinfo{author}{\bibfnamefont{J.~M.} \bibnamefont{{Bardeen}}}, in
  \emph{\bibinfo{booktitle}{Cosmology and Particle Physics}}, edited by
  \bibinfo{editor}{\bibfnamefont{L.-Z.} \bibnamefont{{Fang}}} \bibnamefont{and}
  \bibinfo{editor}{\bibfnamefont{A.}~\bibnamefont{{Zee}}}
  (\bibinfo{publisher}{Gordon and Breach}, \bibinfo{address}{New York},
  \bibinfo{year}{1988}), pp. \bibinfo{pages}{1--64}.

\bibitem[{\citenamefont{Hwang}(1991)}]{hwang91}
\bibinfo{author}{\bibfnamefont{J.-c.} \bibnamefont{Hwang}},
  \bibinfo{journal}{Astrophys. J.} \textbf{\bibinfo{volume}{375}},
  \bibinfo{pages}{443} (\bibinfo{year}{1991}).

\bibitem[{\citenamefont{Komatsu et~al.}(2008)}]{wmap5}
\bibinfo{author}{\bibfnamefont{E.}~\bibnamefont{Komatsu}} \bibnamefont{et~al.}
  (\bibinfo{collaboration}{WMAP}) (\bibinfo{year}{2008}),
  arXiv:0803.0547 [astro-ph].

\bibitem[{\citenamefont{Amendola and Finelli}(2005)}]{af05}
\bibinfo{author}{\bibfnamefont{L.}~\bibnamefont{Amendola}} \bibnamefont{and}
  \bibinfo{author}{\bibfnamefont{F.}~\bibnamefont{Finelli}},
  \bibinfo{journal}{Phys. Rev. Lett.} \textbf{\bibinfo{volume}{94}},
  \bibinfo{pages}{221303} (\bibinfo{year}{2005}), arXiv:astro-ph/0411273.

\bibitem[{\citenamefont{Liddle and Lyth}(2000)}]{ll00}
\bibinfo{author}{\bibfnamefont{A.~R.} \bibnamefont{Liddle}} \bibnamefont{and}
  \bibinfo{author}{\bibfnamefont{D.~H.} \bibnamefont{Lyth}},
  \emph{\bibinfo{title}{Cosmological Inflation and Large-Scale Structure}}
  (\bibinfo{publisher}{Cambridge University Press},
  \bibinfo{address}{Cambridge}, \bibinfo{year}{2000}).

\bibitem[{\citenamefont{Zibin et~al.}(2007)\citenamefont{Zibin, Moss, and
  Scott}}]{zms07}
\bibinfo{author}{\bibfnamefont{J.~P.} \bibnamefont{Zibin}},
  \bibinfo{author}{\bibfnamefont{A.}~\bibnamefont{Moss}}, \bibnamefont{and}
  \bibinfo{author}{\bibfnamefont{D.}~\bibnamefont{Scott}},
  \bibinfo{journal}{Phys. Rev.} \textbf{\bibinfo{volume}{D76}},
  \bibinfo{pages}{123010} (\bibinfo{year}{2007}), arXiv:0706.4482 [astro-ph].

\bibitem[{\citenamefont{Ho et~al.}(2008)\citenamefont{Ho, Hirata, Padmanabhan,
  Seljak, and Bahcall}}]{hhpsb08}
\bibinfo{author}{\bibfnamefont{S.}~\bibnamefont{Ho}},
  \bibinfo{author}{\bibfnamefont{C.~M.} \bibnamefont{Hirata}},
  \bibinfo{author}{\bibfnamefont{N.}~\bibnamefont{Padmanabhan}},
  \bibinfo{author}{\bibfnamefont{U.}~\bibnamefont{Seljak}}, \bibnamefont{and}
  \bibinfo{author}{\bibfnamefont{N.}~\bibnamefont{Bahcall}}
  (\bibinfo{year}{2008}), arXiv:0801.0642 [astro-ph].

\bibitem[{\citenamefont{Mukhanov et~al.}(1992)\citenamefont{Mukhanov, Feldman,
  and Brandenberger}}]{mfb92}
\bibinfo{author}{\bibfnamefont{V.~F.} \bibnamefont{Mukhanov}},
  \bibinfo{author}{\bibfnamefont{H.~A.} \bibnamefont{Feldman}},
  \bibnamefont{and} \bibinfo{author}{\bibfnamefont{R.~H.}
  \bibnamefont{Brandenberger}}, \bibinfo{journal}{Phys. Rept.}
  \textbf{\bibinfo{volume}{215}}, \bibinfo{pages}{203} (\bibinfo{year}{1992}).

\bibitem[{\citenamefont{{Silk}}(1977)}]{silk77}
\bibinfo{author}{\bibfnamefont{J.}~\bibnamefont{{Silk}}},
  \bibinfo{journal}{Astron. Astrophys.} \textbf{\bibinfo{volume}{59}},
  \bibinfo{pages}{53} (\bibinfo{year}{1977}).

\bibitem[{\citenamefont{Hellaby and Krasinski}(2006)}]{hk06}
\bibinfo{author}{\bibfnamefont{C.}~\bibnamefont{Hellaby}} \bibnamefont{and}
  \bibinfo{author}{\bibfnamefont{A.}~\bibnamefont{Krasinski}},
  \bibinfo{journal}{Phys. Rev.} \textbf{\bibinfo{volume}{D73}},
  \bibinfo{pages}{023518} (\bibinfo{year}{2006}), arXiv:gr-qc/0510093.

\bibitem[{\citenamefont{Biswas and Notari}(2007)}]{bn07}
\bibinfo{author}{\bibfnamefont{T.}~\bibnamefont{Biswas}} \bibnamefont{and}
  \bibinfo{author}{\bibfnamefont{A.}~\bibnamefont{Notari}}
  (\bibinfo{year}{2007}), arXiv:astro-ph/0702555.

\end{thebibliography}

\end{document}